\documentclass[fleqn,usenatbib]{mnras}

\usepackage{newtxtext,newtxmath}

\usepackage[T1]{fontenc}
\usepackage{pgfplots}
\usepackage{multirow}
\usepackage[export]{adjustbox}
\DeclareRobustCommand{\VAN}[3]{#2}
\let\VANthebibliography\thebibliography
\def\thebibliography{\DeclareRobustCommand{\VAN}[3]{##3}\VANthebibliography}

\usepackage{subcaption}

\usepackage{graphicx}	
\usepackage{amsmath}	

\usepackage{xspace}

\newcommand{\ha}{H\ensuremath{\alpha}}

\newcommand{\hb}{H\ensuremath{\beta}\xspace}

\newcommand{\hd}{H\ensuremath{\delta_\mathrm{A}}\xspace}

\newcommand{\rb}{\ensuremath{R_{\rm bar}}\xspace}

\title[Bars and Quenching]{The effects of bar strength and kinematics on galaxy evolution II: The global and local impacts of slow-strong bars}

\author[Mengistu et al.]{
Petra Mengistu,$^{1,2}$
Karen L. Masters,$^{1}$
Tobias G\'eron,$^{3,4}$
R. J. Smethurst,$^{3}$
Chris Lintott,$^{3}$
B. D. Simmons$^5$
\\
$^{1}$Departments of Physics and Astronomy, Haverford College,
370 Lancaster Ave.,  Haverford, PA 19041, USA\\
$^{2}$Department of Astronomy and Astrophysics, University of California Santa Cruz,
1156 High St,  Santa Cruz, CA 95062, USA\\
$^{3}$Oxford Astrophysics, Department of Physics, University of Oxford, Denys Wilkinson Building, Keble Road, Oxford, OX1 3RH, UK\\
$^{4}$Dunlap Institute for Astronomy and Astrophysics, University of Toronto, 50 St. George Street, Toronto, ON M5S 3H4, Canada\\
$^5$Department of Physics, Lancaster University, Bailrigg, Lancaster, LA1 4YB, UK
}

\date{Accepted XXX. Received YYY; in original form ZZZ}

\pubyear{\the\year{}}

\begin{document}
\label{firstpage}
\pagerange{\pageref{firstpage}--\pageref{lastpage}}
\maketitle

\begin{abstract}
There is now clear evidence, from a variety of studies, that galactic bars contribute to and/or accelerate processes which quench galaxies. However, bars have a variety of strengths and pattern speeds, and previous work has suggested that slow and strong bars impact their hosts the most. In this paper, we continue to investigate the impact of bar strength and bar speed on host galaxy evolution in a sample of barred galaxies identified via classifications from Galaxy Zoo. We perform a comprehensive assessment of star-formation tracers spanning a variety of timescales, based on spatially resolved spectroscopic information from the Mapping Nearby Galaxies at Apache Point Observatory (MaNGA) survey. Specifically, we examine the radial distributions of EW[\ha\nolinebreak], \hd, \hb, and Dn4000; spectral data that trace star-formation on current, intermediate, and much longer timescales. We investigate how these star-formation tracers vary with respect to each other in diagnostic evolutionary planes for eight categories of barred galaxies (combinations of star forming or quenching; strong and weak; fast and slow). We continue to find that slow-strong bars drive the quenching of their hosts the most by triggering active star-formation throughout the barred region; however, we note some additional complexity: we observe that stronger bars boost star-formation at the bar centre while slower bars have increased star-formation along the bar. This work adds to the growing evidence that galactic bars have both global and local impacts on their host galaxies. 

\end{abstract}

\begin{keywords}
galaxies: evolution  -- galaxies: star formation -- galaxies: bar -- galaxies: kinematics and dynamics -- galaxies: disc 
\end{keywords}



\section{Introduction} \label{sec:intro}
Quenching (or the cessation of star-formation) in galaxies occurs through a variety of mechanisms that are often interdependent (e.g. \citealt{Kormendy2004,Peng2012,Smethurst2017,Bait2017,Bluck2020, Contini2020}). Secular processes (i.e. slow and mostly internal dynamical processes) are understood to be influential in the process of quenching (e.g. \citealt{Kormendy2004,Masters2011}), including in our own Milky Way \citep[e.g.][]{Haywood2016}. In disc galaxies, internal structures such as bars are one of the dominant drivers of secular evolution \citep{Zurita2004, Sheth2005}. 

Bars are rectangular structures in disc galaxies that form as the result of disc instabilities \citep[e.g.][]{Weinberg1985,Debattista2006}. Over time, bars redistribute angular momentum by moving both stars and gas in their host galaxies \citep{Athanassoula2003,Regan2004}. A large fraction of disc galaxies in the present universe contain bars \citep[estimates range from 25-73$\%$ depending on method and wavelength used; ][]{Eskridge2000, Marinova2007,Sheth2008, Nair2010, Masters2011,Buta2019}, and redder, more quiescent, discs are more likely to host bars \citep[particularly strong bars][]{Masters2011,geron2021}, suggesting that bars may play a significant role in disc galaxy quenching.

One of the ways in which bars are believed to impact their hosts is through their ability to redistribute material in the interstellar medium (ISM). Simulations of bars show that they drive radial inflows of gas along dust lanes, accreting gas at the centres of their hosts to form pseudobulges and/or central mass concentrations \citep[e.g.][]{Athanassoula1992, Villa-Vargas2010,Athanassoula2013,Carles2016}. Such inflows can result in rapid star formation enhancement at the centers of barred galaxies \citep[e.g. as observed by ][]{Sheth2000,AlonsoHerrero2001,Ellison2011,Coelho2011,Perez2011,Wang2012,Chown2019,Lin2020}. 

While bars overall tend to be found in discs with suppressed star formation (SF, e.g. as shown in \citealt{Masters2011}), spectroscopic analysis of the SF histories (SFHs) of barred galaxies has revealed that increased SF occurs along the leading edges of bars and bar-ends \citep[e.g. see][]{Peterken2019, DiazGarcia2020,McKelvie2020}. As bars exert a gravitational torque on the surrounding gas, simulations predict that the extent of star formation enhancement is linked to the bar shear potential \citep[e.g. see][]{Athanassoula1992,Renaud2015,Emsellem2015, Khoperskov2018}. Regions with high velocity gradients/shear have suppressed SF whereas regions with lower shear at the bar-ends are associated with triggered SF  \citep{Haywood2016,James2018,Maeda2023, Kim2024, Hogarth2024}. Shocks have also been shown to inhibit SF along the bar-arms \citep{Reynaud1998}, but opposing evidence shows that they may contribute to enhanced SF \citep{Zurita2004}.

Since bars are associated with both enhancement and suppression of star-formation activity, it is no surprise that their overall role in galaxy quenching remains debated. While bar fractions decrease at higher redshifts, these observations of bars suggest that they are long-lived structures that can exist for up to 10 Gyrs in their hosts \citep[e.g.][]{Melvin2014,Guo2025,Geron2025, Salcedo2025}, and so have the potential to act over evolutionary timescales for galaxies. While higher bar fractions are observed in redder, quiescent galaxies relative to those in star-forming galaxies \citep{Masters2011,Masters2012,Cheung2013,Vera2016, Kruk2018,geron2021}, this is sometimes explained by bars being more likely to form in quiescent galaxies rather than themselves contributing to quenching processes.

Not all bars are the same, and the impacts a bar appears to have on its host galaxy depends on its properties \citep[e.g.][]{geron2021}. Strong bars have been suggested to have more observable impacts on their hosts than weak bars \citep{Masters2012,Kim2017,geron2021,geronthesis}, which emphasizes the need to consider bar strength when interpreting the impact of bars on SF activity. Bar strengths \citep[initially a combination of length and brightness;][]{Vaucouleurs1959}, can be estimated by a variety of methods, including analysing the strength of Fourier modes, the bar's contribution to the light distribution profile, assessing the relative size of the bar to the galaxy diameter \citep[e.g.][]{Athanassoula1992,Nair2010,Kruk2018}, or attempting to measure the gravitational torque of the bar \citep{Buta2001,Laurikainen2002} via arm-interam constrast and/or ellipticity of the bar. Large-scale morphological classification of galaxies in recent years has been provided by the Galaxy Zoo (GZ) project \citep{Lintott2008,Willett2013}. GZ obtains information on galaxy morphologies by aggregating classifications collected from online volunteers who are asked to complete a series of tasks identifying morphological features (such as bars) via visual inspection. The GZ ``bar vote fraction" (corrected fraction of users reporting seeing a bar) has been previously used to identify strong and weak bars \citep{Masters2011,Skibba2012}, and furthermore, recent versions of GZ ask for information on bar strength by including classification tasks that asks if a galaxy has a strong, weak, or no bar \citep{geron2021, Walmsley2021,Walmsley2023}, enabling the exploration of the influence of bar strength on their hosts via GZ morphologies.

Bar kinematics is becoming increasingly important in understanding the secular evolution of their hosts, primarily because large resolved spectroscopy surveys enable the measurement of bar pattern speeds, $\Omega_b$, for much larger samples than before \citep[e.g.][]{geron2023}. ``Fast'' bars have been associated with strongly inhibited SF in the arms relative to ``slow'' bars, indicating that bar speed may also influence SF within the barred region \citep{geronthesis}. The length of the bar relative to its corotation radius determines the speed at which the bar moves relative to stars at the bar end. To quantify this, the dimensionless quantity $\mathcal{R}=R_{\mathrm{CR}}/R_\mathrm{{bar}}$ is defined as the ratio of the corotation radius to bar radius. The {corotation radius} of a bar, $R_{\mathrm{CR}}$ is the radius at which stars in the galaxy have the same angular speed as the bar (stars interior to this radius move faster, and those outside move slower). ``Fast'' bars end at or near co-rotation ($1.0<\mathcal{R}<1.4$) so their ends move with the stars in the disc, while ``slow bars'' end inside co-rotation ($\mathcal{R}>1.4$; and thus rotate more slowly than the {disc} stars at their ends, \citealt{Tremaine1984, Elmegreen1996}).  

It is important to consider the integrated influence of bar strength and kinematics to understand the processes by which bars drive the evolution of their hosts. Recent work by \cite{geronthesis} probed the impact of bar strength and bar speed on SF in different types of barred galaxies at different timescales in the host's evolution by examining the radial distributions of EW[\ha] (a tracer of current SF) and Dn4000 (a tracer of SF beyond ~Gyr timescales) along the barred region. Among various subsets of barred galaxies, split based on combinations of host SFR, bar speed, and bar strength, they observe that young SF galaxies with bars that are both strong and slow have increased ongoing SF and younger stellar populations within the bar than other types of barred galaxies; additionally, \cite{geronthesis} find that this increase in SF is local to the bar-end region of galaxies with slow and strong bars compared to other surrounding regions in the hosts.

This, along with other observations pointing to secular evolution driven by bars \citep[e.g.][]{Cheung2013,Gavazzi2015, Renu2025,deSaFreitas2025}, suggests that bars are influential contributors to morphological quenching. Both theoretical insights and observations suggest that bars slow down \citep{Algorry2017,Cuomo2021,geron2023} and grow longer (and therefore more likely stronger) over time \citep{Villa-Vargas2010,Kim2017,McKelvie2020,geron2021, Rosas-Guevara2021, Fragkoudi2025, deSaFreitas2025}. \cite{geronthesis} propose that bars progress along a continuum of strengths and speeds and therefore their impact on their hosts increases according to their evolutionary stage, such that galaxies with slower and stronger bars are the most quenched.

In this paper, we build on the work presented in \cite{geronthesis} by investigating how bar speed and bar strength interact through detailed examinations of the SFHs of various barred galaxies using the tracers EW[\ha] and Dn4000 previously considered in \cite{geronthesis} that probe current and distant timescales, but also adding two additional tracers (\hb and \hd) that probe SF on intermediate timescales. By adding these tracers, we are better able to examine the trajectory of bars of different strengths and speeds in SF and quenching host galaxies across evolutionary planes (i.e. EW[\ha], \hb, and \hd vs. Dn4000) which diagnose star formation histories. Additionally, we assess the extent of the bar's influence by comparing the impact of bar strength/speed on localized turnovers of SF within the bar region to those of global galaxy properties, and consider how the \ha~morphology helps us to understand the trends we see.

The rest of this paper is structured as follows: in Section \ref{sec:methods}, we describe the data and methods used to analyse star-formation across the barred region; in Section \ref{sec:results}, we present the radial profiles of the four selected spectroscopic tracers of star-formation (EW[\ha], \hb, \hd, Dn4000) in various subsets of barred galaxies and analyse the correlations between these measures of star-formation, bar strength, bar speed, and other host physical properties; in Section \ref{sec:discussion}, we discuss these results and present our conclusions in Section \ref{sec:conclusion}.

The only distance dependent quantity used in this work is stellar mass from Pipe3D \citep[][see Section \ref{sec:sample}]{Sanchez2016} which is used to split the sample into SF and quenching galaxies based on the SF sequence of galaxies. Pipe3D is a spectral fitting algorithm applied to MaNGA data (see Section \ref{subsec:Manga} below) and assumes standard $\lambda$CDM with $H_0 = 73$km/s/Mpc to derive distant dependent quantities. \citep{Sanchez2022}.

\section{Data and Sample Selection} \label{sec:methods}
\subsection{Galaxy Zoo and Identification of Strong and Weak Bars}\label{subsec:gz}
We use Galaxy Zoo (GZ, \citealt{Lintott2008}), a crowd-sourced citizen science project that provides morphological classifications for large datasets of galaxies, to obtain our sample of barred galaxies and to identify strong and weak bars. Recent versions of GZ ask about both strong and weak bars, and have developed machine learning algorithms trained on participant responses from prior campaigns to make morphological classifications for larger samples \citep{Walmsley2021, Walmsley2023}. This paper makes use of GZ classifications of the Dark Energy Spectroscopic Instrument \citep[DESI, ][]{Dey2019} survey images. GZ DESI provides machine-generated classifications for approximately 8.7M galaxies, trained on GZ volunteer classifications \citep{Walmsley2023}. We base our sample on a set of barred galaxies selected from GZ DESI by \citet[][see Section 2.5 below for further details on sample selection]{geron2023}.

In order to extract samples with reliable classifications from GZ, thresholds must be imposed on the vote fractions, or the corrected fraction of the participants who reported seeing the feature, labeled as ``p[primary response]" \citep[e.g. see ][ for details on this]{Willett2013}. We use the classifications from GZ DESI, which included a question that asks whether a galaxy has a strong, weak, or no bar \citep{Walmsley2023}. We impose the restrictions on the subsequent vote fractions ($p_\mathrm{{strong}}, p_\mathrm{{weak}}, p_\mathrm{no \,bar}$) defined in \cite{geron2021} to identify strong and weak bars. A galaxy is considered to host:
\begin{itemize}
    \item a strong bar:  if at least half of the classifications report that a bar is present ($p_\mathrm{{strong+weak}} \ge 0.5$) and there are more classifications as a strong bar than weak ($p_\mathrm{{strong}} \ge p_\mathrm{{weak}}$)
    \item a weak bar:  if $p_\mathrm{{strong+weak}} \ge 0.5$ and there are more classifications as weak than strong ($p_\mathrm{{strong}} \le p_\mathrm{{weak}}$). 
    \item no bar: if total bar vote fraction,  $p_\mathrm{{strong+weak}} \le 0.5$.
\end{itemize}

\label{subsec:Manga}
\subsection{MaNGA}
MaNGA (Mapping Nearby Galaxies at Apache Point Observatory) was a survey of nearby galaxies forming part of SDSS-IV \citep{Bundy2015, Blanton2017}. Over a six year period (2014--2020), MaNGA observed 10,001 unique galaxies with the MaNGA integral field unit (IFU) and the BOSS spectrograph \citep{Smee2013,Drory2015} attached to the 2.5m Sloan Foundation Telescope \citep{Gunn2006}. MaNGA provides spatially resolved spectral measurements for the entire coverage of each galaxy. This typically extends to $\sim$ 1.5 $R_e$ but can extend up to 2.5 $R_e$, where $R_e$ is effective radius \citep{Wake2017}. Each IFU is composed of optical fibre bundles arranged in a hexagonal array that collectively provide 17--32" of coverage, with effective resolution of 2.5". The reduced data are made publicly available in the form of the MaNGA Data Reduction Pipeline (DRP; see \cite{Law2016}; subsequent analysis of this data is available as the MaNGA Data Analysis Pipeline \citep[DAP][]{Westfall2019}. To access MaNGA data, we make use of the {\it Marvin} software \citep{Cherinka2019}. 

The MaNGA DAP provides two-dimensional maps of various emission and absorption features. We use the measurements of four spectral features that indicate star-formation activity at various timescales. Specifically, we use measurements of:
\begin{itemize}
    \item \textbf{Gaussian-fitted Equivalent Width of H$\alpha$ (EW[\ha])} which traces the flux of \ha~ emission (located at 6564 \AA), indicating ongoing star formation within a galaxy.
    
    \item \textbf{H$\beta$ absorption} index which probes the immediate, or very recent, cessation of star formation \citep[e.g. see][]{Smethurst2019}. The absorption of light by hydrogen in stellar atmospheres at this wavelength (4861$\rm{\mathring{A}}$) is only visible in the absence of OB stars. 
   
    \item \textbf{H$\delta$ absorption} feature which also serves as a tracer of recent, rapid quenching in galaxies. As the absorption of light at this wavelength (between 4080\AA -- 4122\AA) is dominated by A type stars, H$\delta$ is most sensitive to star-formation activity within the last $\sim$ Gyr \citep{ Balogh1999,Goto2003, Smethurst2019}.
   
    \item \textbf{D$_\mathrm{n}$4000}, the ratio of the total flux between 4000-4100\AA ~and 3850-3950\AA ~wavelengths, which traces the presence of older stellar populations. While this correlates with the stellar population age, it is important to consider that there is a distinct metallicity dependence for stellar ages above 1 Gyr \citep{Kauffmann2003}.
\end{itemize}
We apply corrections to the spectral index measurements (\hb, \hd, and Dn4000) as advised in \cite{Westfall2019,Cherinka2019} to account for the stellar and gas velocity dispersions due to constraints on instrument resolution. For measures of EW[\ha], we ensure that we only trace star-formation activity and avoid contamination by other ionization sources by removing spaxels which lie in the Active Galactic Nuclei (AGN), Low Ionization Nuclear Emission line-Regions (LINER), Seyfert, and composite star-forming regions on a Baldwin, Phillips, \& Terlevich (BPT) diagram \citep[][using the BPT function in {\it Marvin}]{Baldwin1981}. 

We additionally use stellar masses based on the spectral fitting from data integrated across the MaNGA bundle from the Pipe3D value-added-catalogue (VAC) for MaNGA \citep{Sanchez2016,Sanchez2022}.

\subsubsection{Bar Speeds}
\citet{geron2023} make use of MaNGA kinematic data to obtain measurements of the bar kinematics, which we use in this work. Using the bar corotation radius $R_{\rm CR}$ and the deprojected bar radius $R_{\rm bar}$, the usual measure of bar speed is given by 
\begin{equation}
    \mathcal{R} = \frac{R_{\rm CR}}{R_{\rm bar}}.
\end{equation}
Fast bars are defined to have $1.0<\mathcal{R}<1.4$ and slow bars have $\mathcal{R}>1.4$. To obtain $\mathcal{R}$, measures of the bar pattern speed, galaxy rotation curve (to obtain the co-rotation radius), and bar radius are required. We use the measures of deprojected bar radius, bar pattern speed, and $\mathcal{R}$ provided by \cite{geron2023}\footnote{all of which are publicly accessible here  \href{https://zenodo.org/record/7567945}{https://zenodo.org/record/7567945}}, and will summarize the method they use below. 

Various methods can be used to obtain estimates of bar pattern speeds
\citep[e.g.][]{Canzian1993,Rautiainen2008,Treuthardt2008,Font2011}. However, the most widely used, and only model-independent method is the \citet[][TW]{Tremaine1984} method. The TW method estimates the bar pattern speed using kinematic tracers such as stars or gas which must be assumed to satisfy the continuity equation.

For more details on this method, see \citet{geron2023}, or for a full derivation from the continuity equation, see \cite{Binney&Tremaine2008}.
While the assumption of continuity holds well when applied to stars in quiescent galaxies \citep{Merrifield1995}, there is ongoing discussion on whether gas, or stars, in SF galaxies can be used as suitable tracers. In these galaxies, active SF and dust obscuration may limit the ability to fully trace the underlying mass distribution \citep{Tremaine1984, Gerssen2007, Aguerri2015}. Despite these potential limitations, the TW method has been applied to both quiescent and SF galaxies using stars as tracers \citep{Cuomo2019, Guo2019,GarmaOehmichenn2020,Cuomo2020, Cuomo2021, Williams2021, GarmaOehmichen2022}. 

\cite{geron2023} obtain bar pattern speeds using spatially resolved fluxes and line-of-sight velocities from MaNGA for barred galaxies. Luminosity weighted averages of the position and line-of-sight velocity (corresponding to the photometric and kinematic integrals respectively) are obtained by summing over 0.5" wide pseudoslits that are centered on the disc minor axis, aligned with the galaxy major axis, and span the integral field unit's field of view. \citet{geron2021,geron2023}  provide measured bar lengths based on visual inspection (by TG) of $grz$ optical images from the DESI Legacy surveys.

Finally, the corotation radius of a bar (the distance at which gravitational force and centrifugal forces are balanced in the bar rest frame) is the radius at which the rotation curve intersects with the bar's pattern speed expressed as a rotational velocity ($\Omega_b r$). To obtain rotation curves, \cite{geron2023} fit MaNGA stellar velocities to a two-parameter arctan function defined in \cite{Courteau1997}:
\begin{equation}
    V_{\rm rot} = V_{\rm sys} + \frac{2}{\pi} V_c \, \, \mathrm{arctan} \left( \frac{r - r_0}{r_t}\right)
\end{equation}
where $V_{\rm rot}, V_{\rm sys}$ and $V_c$ are the rotational, systemic, and circular velocities respectively; $r$ is the deprojected distance from the galaxy center, $r_0$ is the position of the galaxy center, and $r_t$ is the transition radius.

\subsubsection{HI-MaNGA}
To obtain measures of the HI content of galaxies, we use data from DR4\footnote{https://greenbankobservatory.org/portal/gbt/gbt-legacy-archive/hi-manga-data/} of the HI-MaNGA programme \citep{Masters2019,Stark2021}. HI-MaNGA obtains HI 21cm radio observations using the L-band receiver on the Robert C. Byrd Green Bank Telescope (GBT), and also provides consistent information on HI observations of MaNGA galaxies available from other surveys (mostly from ALFALFA,  \citealt{Haynes2018}). 
HI-MaNGA provides information on both HI detections and upper limits for MaNGA galaxies $z<0.05$. HI masses for detections are measured via
\begin{equation}
    M_{\rm HI}/M_\odot = 2.536 \times10^5 \left( \frac{D}{\mathrm{Mpc}}\right)^2 \left(\frac{F_{\rm HI}}{\mathrm{Jy \, km \, s^{-1}}} \right)
\end{equation}
where $M_{\rm HI}$ is the HI mass, $D$, is the distance to the galaxy, and $F_{\rm HI}$ is the observed HI flux). For non-detections, an upper limit of the HI mass is provided using the $rms$ noise of the spectrum and assuming a width of $W=200 ~\mathrm{km \, s^{-1}}$ to estimate the 1$\sigma$ upper limit.  

Using HI mass fractions ($f_{\rm HI} = M_{\rm HI}/{M_\star}$), the HI deficiency $\rm{HI}_{\rm def}$ is quantified by taking the difference between the expected gas fraction $f_{\rm HI_{exp}}$ and the observed HI mass fraction, or 
\begin{equation}
    \label{eqn:HIdef}
    \mathrm{HI}_{\mathrm{def}}=f_{\mathrm{HI_{exp}}} - \log \left( \frac{M_{\mathrm{HI}}}{M_*} \right). 
\end{equation}
We follow \citet{Masters2012} for the trend of  $f_{\rm HI_{exp}}$ with stellar mass assuming
\begin{equation}
    \label{eqn:HIdef_exp}
    f_{\mathrm{HI_{exp}}} = -0.31 - 0.86\left(\log \left( \frac{M_{\star}}{M_\odot} \right) - 10.2\right).
\end{equation}

\subsubsection{Galaxy Environments for MaNGA (GEMA)}
To probe the impacts of galaxy interactions we make use of measures of tidal strength from the Galaxy Environments for MaNGA (GEMA) Value-Added Catalog released as part of SDSS-IV DR17 \citep{Abdurro'uf_sdssdr17}. GEMA provides various mass-dependent (tidal strength $Q$) and mass-independent (projected density $\eta$) measures of galaxy environment, quantified following the methods in \cite{Argudo2015}. The tidal strength parameter estimates the net strength of the gravitational interaction exerted on the primary galaxy due to binding forces with its neighbors, defined by \cite{Argudo2015} as
\begin{equation}
    Q \equiv \mathrm{log} \left( \sum_i \frac{M_i}{M_P} \left( \frac{D_P}{d_i}\right)^3 \right)
\end{equation}
where $M_P$ is stellar mass of the primary galaxy, $M_i$ is the stellar mass of the $i$th neighbor galaxy, $D_p$ is the estimated diameter of the primary galaxy corresponding to the Petrosian radius at which 90$\%$ of the flux is contained, and $d_i$ is the distance from the $i$th neighbor to the primary galaxy. As the sample we use is volume-limited by $z<0.05$, we use the measure of tidal strength of the first nearest neighbor provided in GEMA for neighbors within $500 \mathrm{km \, s^{-1}}$ line-of-sight velocity, volume-limited by $z<0.06$, and projected up to a distance of $5 \,\rm{Mpc}$. 

\subsection{Sample Selection}\label{sec:sample}
To investigate the impacts of bar strength and bar kinematics on galaxies, we use the sample (hereafter referred to as the Tremaine-Weinberg, or "TW" sample) presented in \cite{geron2023} which consists of barred galaxies with bar pattern speeds measured using the TW method. We provide a summary of the selection criteria \citet{geron2023} used to obtain this sample here: 
\begin{enumerate}
    \item Galaxies with both morphological and resolved spectral information are obtained from a cross-match of the GZ DESI dataset ($\sim$ 8.7 million galaxies) and the MaNGA survey ($\sim$ 10,001) galaxies, resulting in an initial sample of 9,817. 
    \item Magnitude and redshift limits of 0.01 $< z <$ 0.05 and $M_r<-18.96$ are applied  \citep[both of which are obtained from the NASA Sloan Atlas;][]{Blanton2011}, reducing the number of galaxies to 5,810. 
    \item Threshold requirements are imposed on the predicted vote fractions from GZ DESI to reliably select galaxies that have distinctly observable features ($p_{\rm features}>0.27$), that those features are highly likely to be barred face-on spirals ($p_{\rm notedge\_on}>0.68$), and the estimated number of volunteers that would make the bar classification ($n_{\rm bar}>0.5$). This reduces the number of galaxies to 2,125 candidate face-on barred galaxies. 
    \item Of the galaxies with potentially observable bars, 1,213 galaxies have bar lengths and position angles measured by \cite{geronthesis}. 
    \item To measure bar pattern speeds using the TW method, face-on barred galaxies that have regular kinematics are selected; see \cite{geron2023, geronthesis} for a full review), which reduces the sample size to 210 galaxies with measured bar pattern speeds ($\Omega_b$), corotation radii (R$_{\mathrm{CR}}$), and $\mathcal{R}$. 
   \end{enumerate} 
    We use the TW sample to examine the impact of bar speed on radial distributions of star-formation indicators relative to bar strength and other global galaxy properties.

We identify galaxies within the TW sample that are SF and quenching using the star-forming sequence defined by \cite{Belfiore2018} for star-formation rates (SFRs) and stellar masses derived from the Pipe3D pipeline \citep{Sanchez2016}. 
\begin{equation}
    \label{eq:sfs}
    \mathrm{log} (\mathrm{SFR/M_\odot yr^{-1}}) = (0.73 \pm 0.03)\,\mathrm{log}(\mathrm{M_*/M_\odot}) - (7.33 \pm 0.29)
\end{equation}
Galaxies with SFRs and masses below 1$\sigma (\sim 0.39$ dex) scatter of the star-forming sequence in Equation \ref{eq:sfs} are classified as quenching and the rest as star-forming. 

Of the 210 galaxies in the TW sample (see Section \ref{sec:sample}), 196 have been observed by the HI-MaNGA programme and 123 have identified detections from the fourth data release (DR4). We determine the HI deficiency for the subset of 123 galaxies with HI detections, which we call the HI rich subset. There are well-known correlations between HI detections and the presence of a (strong) bar \citep{Masters2012}, and in addition this selection preferentially selects more SF galaxies due to the correlations between SF and HI \citep[especially long-term tracers of SF, as shown with HI-MaNGA data in ][]{Stark2021}. We find that out of the 73 barred galaxies that are HI upper limits in HI-MaNGA, galaxies with slow bars are also slightly more likely to be upper-limits than fast bars. Specifically, we find 52/140, or 37\% of slow bars are upper-limits, while that same fraction is 21/70, or 30\% of fast bars. This means the HI rich subset is slightly biased against galaxies with strong and/or slow bars.


A summary of the final sample sizes is provided in Table \ref{tab:samples}. We note that, even starting from the 10,001 MaNGA galaxies, our smallest subset (SF galaxies with fast-weak bars) contains only 10 examples.

\begin{table}
\centering
\caption{\label{tab:samples} A summary of all the sizes of all subsamples of the total 210 barred galaxies in our study. SF=Star forming, Q=Quenching.} 
\begin{tabular}{cc|c|c|c}
\multicolumn{1}{c|}{}    &    & Fast & Slow & Fast + Slow  \\ \hline \hline \hline
\multicolumn{1}{c|}{\multirow{2}{*}{Strong}}              & SF & 18   & 26   & 44                  \\ 
\multicolumn{1}{c|}{}                                     & Q  & 29   & 40   & 69                  \\ 
\hline
\multicolumn{1}{c|}{}                                     & All  & 47   & 66   & 113               \\ 
\hline
\hline
\multicolumn{1}{c|}{\multirow{2}{*}{Weak}}                & SF & 10   & 37   & 47               \\
\multicolumn{1}{c|}{}                                     & Q  & 13   & 37   & 50                \\ 
\hline
\multicolumn{1}{c|}{}                                     & All  & 23   & 74   & 97               \\ 
\hline
\hline
\multicolumn{1}{c|}{\multirow{2}{*}{Strong+Weak}} & SF & 28   & 63   & 91               \\
\multicolumn{1}{c|}{}                                     & Q  & 42   & 77   & 119        \\ 
\hline
\multicolumn{1}{c|}{}                              & All       &  70 & 140    & 210 \\ 
\hline \hline  \hline                
\end{tabular}
\end{table}

\section{Results} \label{sec:results}
\begin{figure}
    \centering
    \includegraphics[width=\linewidth]{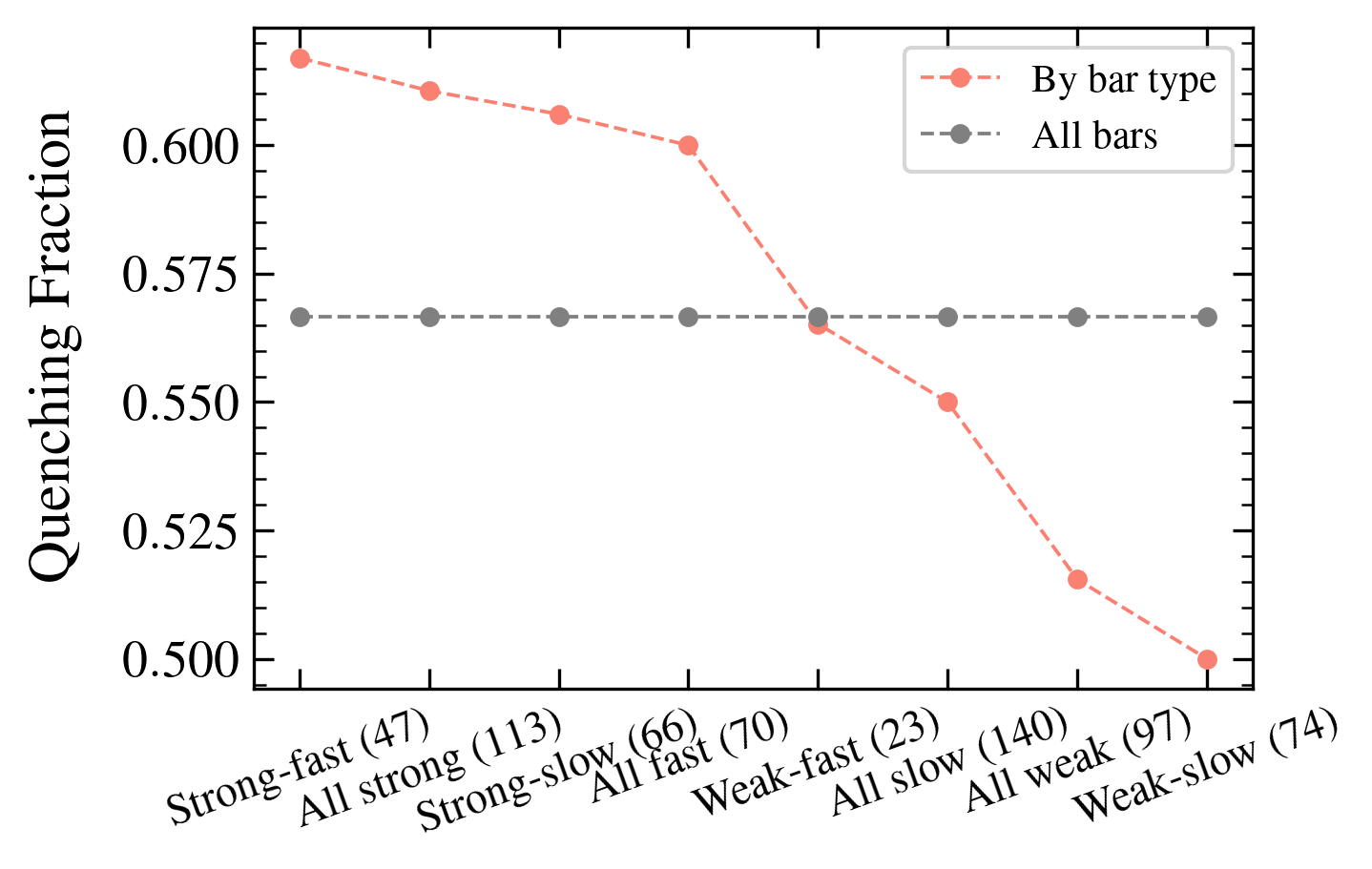}
    \caption{{Fraction of galaxies that are quenching in each bar-type of the TW sample (peach line) and of the total TW sample shown for comparison (gray dashed line). We note that the quenching fractions by bar type are not statistically different than that of the entire population within the Poisson errors for this sample size (which are presented in the Section \ref{sec:results}), but the trends suggest that more galaxies with strong bars are quenching than those with weak bars, and more with fast bars than slow.}}
    \label{fig:quenching_fractions}
\end{figure}

{We first summarize the distribution of bar and host galaxy properties across the TW sample. We note that this is as previously reported in \cite{geronthesis} and provide the following to help the reader interpret our additions to that work}. In the 210 barred galaxies with both strength and speed measurements, we note that barred galaxies are slightly more likely to be quenching (56$\pm$5\% of barred galaxies are quenching) than SF (43$\pm$5\% are SF; all sub-sample sizes are summarized in Table \ref{tab:samples}). At face-value, the TW sample is close to 50-50 quenching-SF, but recall that disc galaxies overall are preferentially SF galaxies, so this balance represents a bias of barred galaxies towards being quenching. {Figure \ref{fig:quenching_fractions} shows the fractions of quenching galaxies in each of the different bar types (peach line) - while these fractions are not statistically different from that of the entire population, the overall trends show that more strongly barred galaxies are quenching relative to their weaker counterparts; secondarily, more galaxies with fast bars are quenching relative to their slower counterparts.} 

A total of 54$\pm$5\% of the bars in our sample with TW measurements are strong bars, with the remainder 46\%$\pm$5\% being weak. The TW measurements reveal that two-thirds of bars in our sample (67$\pm$6\%) are slow bars and the remaining third (33$\pm$4\%) are fast bars, as also reported in \cite{geronthesis}. {We note that a majority of bars in the TW sample are slow; however, as described in \cite{geron2023}, this may be an impact of {shorter bar lengths being more likely to have complete MaNGA bundle coverage} rather than slower rotation speeds. Shorter bars have larger $\mathcal{R}$ and are therefore slower, which \cite{Frankel2022_short_not_slow} suggest may explain the overabundance of slow bars in simulations. As weak bars (which are shorter by definition) are included in the TW sample, an abundance of shorter bars may explain the observed skewed distribution of $\mathcal{R}$ towards larger (slower) values.} 

Within these barred galaxies, we observe that galaxies with slow or fast bars are about equally likely to be quenching (55$\pm$6\% compared to 60$\pm$9\%); however, galaxies with strong bars are slightly more likely to be quenching compared to galaxies with weak bars (61$\pm$7\% strong bars are quenching compared to 52$\pm$7\% weak bars). Strong bars are also less likely to be slow bars: 58$\pm$7\% of strong bars are slow, compared to 76$\pm$9\% of weak bars. 

 With these sample distributions in mind, we now move on to investigating the radial profiles of SF indicators. 

\begin{figure*}
    \centering
    \includegraphics[width=\textwidth]{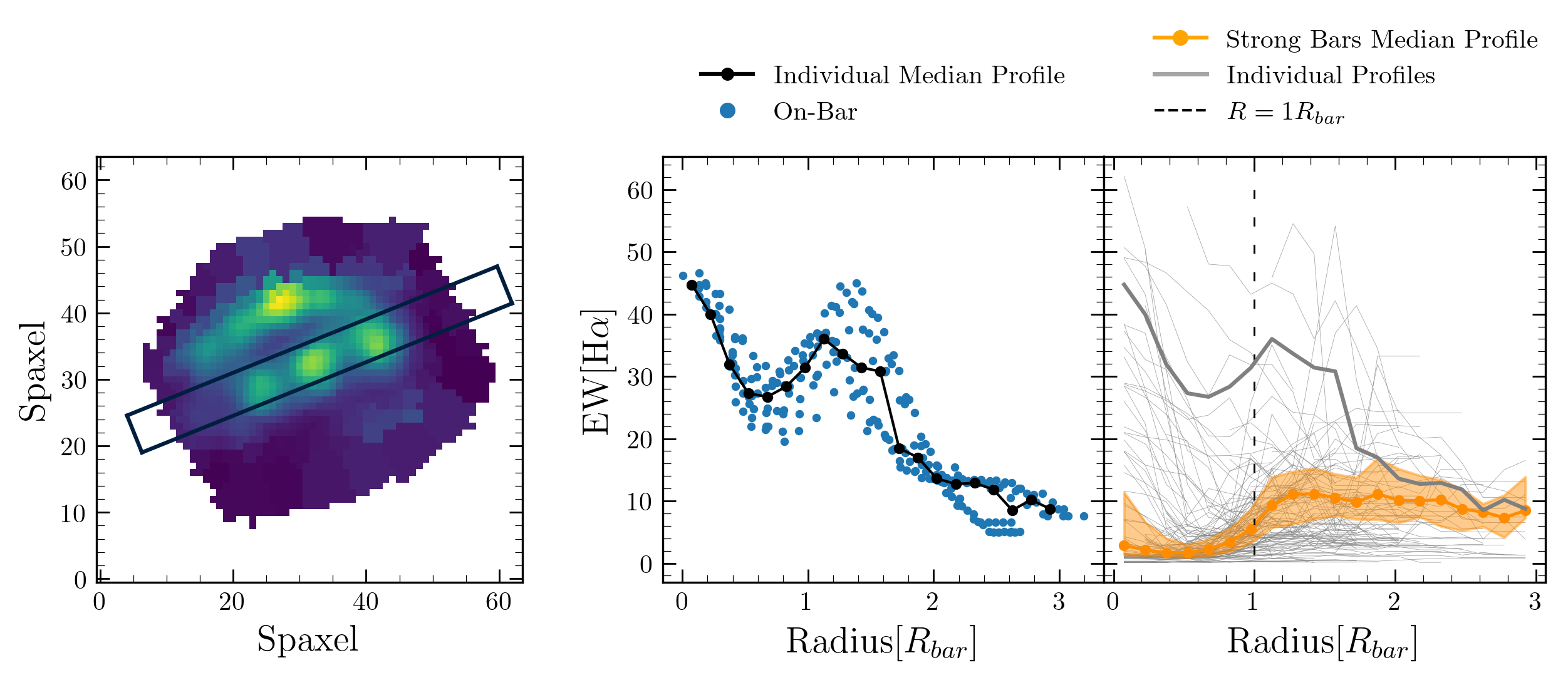}
    \caption{An illustration the process of constructing radial profiles, {similar to that shown} in Figure 2 in \citet{geronthesis}. The left panel shows the EW[\ha] distribution with a 3"x30" aperture aligned with the position angle of the bar for the MaNGA galaxy, \texttt{plate-ifu=12622-9101} \citep[the same example used in][]{geronthesis}. The middle panel shows EW[\ha] plotted against the distance from the galaxy center normalized to the bar radius for the region with spaxels whose centers lie within the aperture. A median profile for this galaxy (shown in black) is obtained by taking the median EW[\ha] within bins of size 0.15 \rb. {This differs slightly from that shown in \citet{geronthesis} due to slightly different choices about data corrections}. The rightmost panel shows the total median profile obtained by taking the median of all individual profiles within the strongly barred subset of the TW sample (113 galaxies), with the shaded orange regions spanning the 33rd and 66th percentile. We show the contributing individual median profiles in light gray, with the profile of the example MaNGA galaxy shown in a thicker gray line.  {This plot is for the TW sample that we solely use in this work \citep[and is one of the samples used in][]{geronthesis}, whereas Figure 2 in \citet{geronthesis} shows this for the DESI-MaNGA sample \citep[another sample used in][of which the TW sample is a subset]{geronthesis}}. }
    \label{fig:prof_ex}
\end{figure*}

\subsection{Radial Profiles} 
\label{sec:rad_prof}
We investigate how star-formation varies spatially across the barred region for each of our barred galaxy subsets using four spectroscopic tracers of star-formation (EW[\ha], \hb, \hd, and Dn4000). We examine the radial distributions of these star-formation tracers across regions aligned with the bar position angle. To generate radial profiles of these tracers, we follow the procedure in \cite{geronthesis} {who reported similar radial profiles, but only for EW[\ha]\footnote{{Additionally \cite{geronthesis} applied different corrections to EW[\ha] from those we describe in Section 2.2, so our values differ slightly.}} and Dn4000; we add additional profiles for new tracers, and consider some additional and/or different subgroups by bar properties. The \citet{geronthesis} procedure was to} overlay an aperture\footnote{{We define our own apertures in this work, although we note that it is possible to use the \texttt{getAperture} method available in \textit{Marvin} \citep{Cherinka2019}. We have confirmed that using either method yields consistent results.}} of size 3" x {30"} onto the MaNGA maps at the PA of the bar (see Figure \ref{fig:prof_ex}). The choice of the dimensions for the aperture are somewhat arbitrary; however, they both cover the resolution (typically 2.5\arcsec) and field of view of the MaNGA data, and by visual inspection, \cite{geronthesis} determined that they capture the majority of most bars. The left panel of Figure \ref{fig:prof_ex} shows this aperture on an example galaxy.

For every spaxel\footnote{In MaNGA data, a spaxel is a spatial pixel, of size 0.5" to enable Nyquist sampling of the $\sim$2.5" effective resolution \citep{Law2016}} whose centre lies within the aperture, we extract the value of the star-formation tracer and determine the distance from the given spaxel to the centre of the galaxy, normalized to the bar radius. We then derive the individual median profile for a galaxy by plotting median values in bins of size of 0.15 R$_\mathrm{{bar}}$ (the bar radius) against the normalized distance from the galaxy centre (shown in the central panel of Figure \ref{fig:prof_ex}). To obtain total median profiles representative of all galaxies within a given subset, we construct the individual median profiles for each galaxy using the process described above and take the median of all the individual profiles as all bins are equally sized (rightmost panel of Figure \ref{fig:prof_ex}).

We show the radial profiles of SF and quenching galaxies hosting all subsets of fast/slow and strong/weak bars in Figure \ref{fig:sf_combo}. {We reiterate that while similar radial profiles of SF galaxies for the EW[\ha] and Dn4000 tracers were previously shown in \cite{geronthesis}, we add here profiles for quenching barred galaxies and add the \hb~and \hd tracers in all galaxies in the TW sub-samples. Additionally we perform slightly different corrections to the values of these measures (see Section 2.2). We show all profiles here for ease of comparison.} The profiles of SF galaxies are shown in the left panels and those of quenching galaxies on the right, with fast-strong bars shown in purple, slow-strong in green, fast-weak in pink, and slow-weak in blue. The main panels are ordered top-to-bottom by timescale of the star-formation tracer shown (i.e. EW[\ha] -- \hb -- \hd -- Dn4000) and the residuals between profiles of barred galaxies by bar speed (i.e. slow-strong -- fast-strong and slow-weak -- fast-weak) are shown in each sub-panel, with differences greater than $3\sigma$ highlighted with red points. The vertical dashed line in all panels indicates the length of the bar.

Overall, we observe {the same qualitative evidence} from the radial profiles that where SF is ongoing, \textbf{slow-strong bars trigger the most star-formation in their still SF hosts} {as reported in \cite{geronthesis}}; {we note that we observe additional support for these trends in the new profiles of the two post-starbust tracers (\hb and \hd) in addition to those of \ha~and Dn4000 shown in \cite{geronthesis}.} 

{Additionally, we observe new trends in the profiles of quenching galaxies that show that} {\bf those hosting slow-strong bars are the most quenched.} {We also observe new trends that galaxies with slow-weak bars are highly passive when in the quenching phase but, as seen in \cite{geronthesis}, do not show signatures of enhanced star-formation in star-forming galaxies}.

Galaxies with {\bf fast bars} (both strong and weak - purple and pink lines) {\bf have both suppressed star-formation in the bar itself and enhanced star-formation at the bar-ends}. This is present in both SF {\citep[as noted by][]{geronthesis}} and globally quenching galaxies ({which we report for the first time in this work}). SF galaxies with {\bf fast-strong bars have the most suppressed SF in the bar region} (purple lines). This implies fast bars (weak or strong) are inhibited from star-formation within the bar of their hosts, even if significant gas is present. {We observe new evidence that} quenching galaxies with fast-weak bars have the most ongoing SF at the bar-ends. We will expand on these observations in the sub-sections below.

\subsubsection{Star-forming Barred Galaxies}
\label{sec:sf_combo}

\begin{figure*}
     \includegraphics[width=0.78\textwidth]{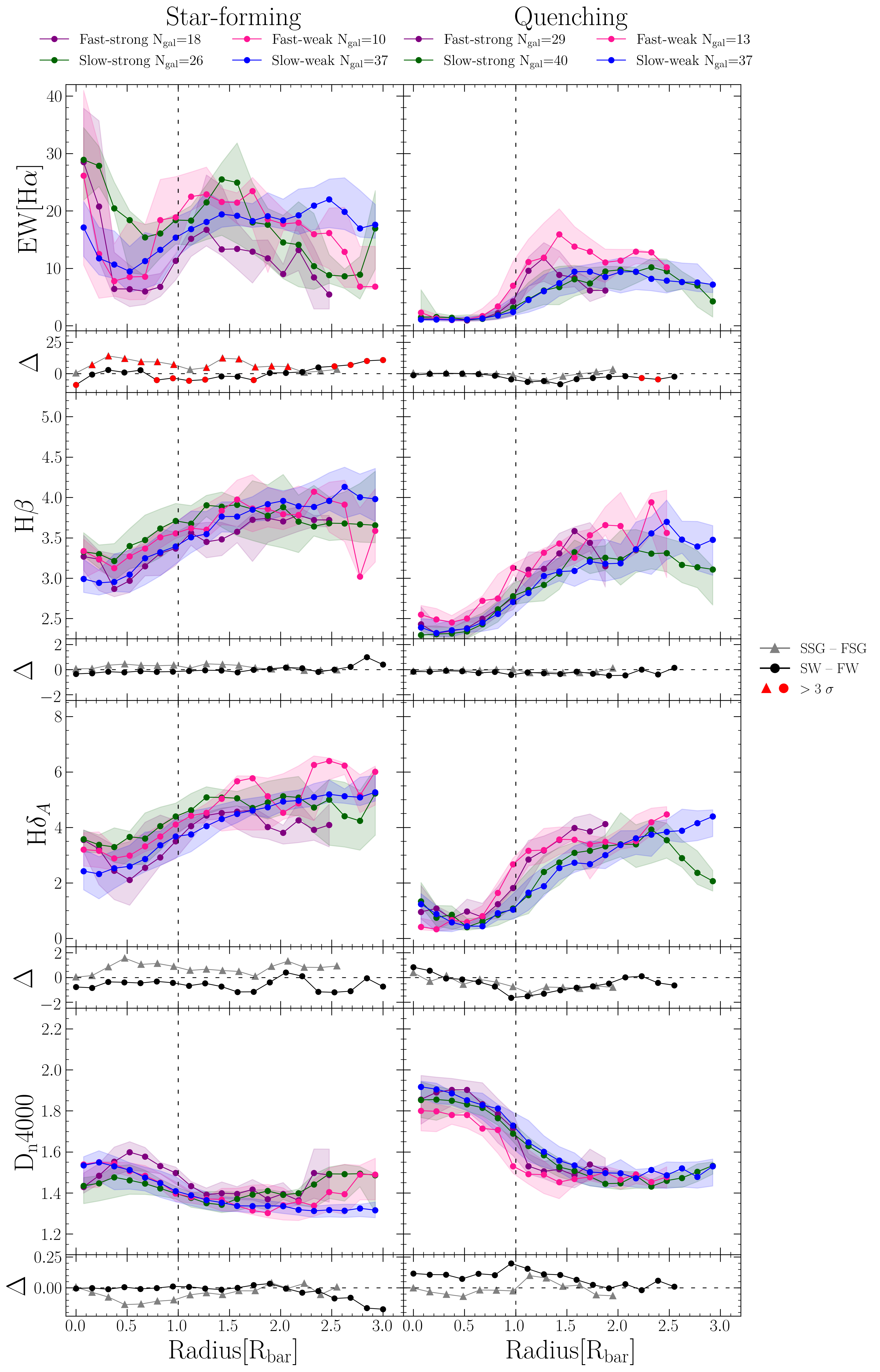}
     \caption{Radial profiles of EW[\ha], \hd, \hb, and Dn4000 for star forming galaxies (left panels) and quenching galaxies (right panels) sub-divided by both bar strength and speed (main panels). Shaded regions show the 33rd-66th percentile and the points are median values within bins of size 0.15\rb. Smaller sub-panels beneath each main panel show the difference between the profiles of bars of the same strength but of different speeds in the corresponding main panel (i.e. Slow/strong--Fast/strong and Slow/weak--Fast/weak, represented by the acronyms SSG, FSG, SW, and FW respectively). Where the residuals between the profile pairs are statistically significant ($>3 \sigma$) using an Anderson-Darling test, points are colored red. Slow-strong bars in SF host galaxies have higher sSFRs, younger stellar populations, and more recent/recently completed SF within the barred region.}
     \label{fig:sf_combo}
\end{figure*}

The EW[H$\alpha$] profiles (top-left panel of Figure \ref{fig:sf_combo}) of all star forming barred galaxies show a central peak, dip in the bar region, and increase beyond the end of the bar. This is mirrored in the Dn4000 profiles (lower-left), which reveal younger central stellar populations, a peak in the bar region and decline beyond the bar, as previously seen in \cite{geronthesis}. Both \hd and \hb profiles for SF galaxies show subtle central peaks and dips in the bar region, then a monotonic increase from $R\sim R_{\rm bar}/2$ to the outskirts.  

There are notable differences between SF galaxies with bars of different strengths and speeds. 
\begin{itemize}
    \item {\bf Star forming galaxies with Slow-strong bars} (green lines in Figure \ref{fig:sf_combo}) show the highest levels of EW[H$\alpha$] throughout the barred region, peaking at the bar centre and {just beyond the bar-}end \citep[as also observed in][]{geronthesis}. This is a statistically significant difference ($>3\sigma$ by the Anderson-Darling test) between slow-strong (SSG) and fast-strong (FSG) bars.  This trend is mirrored in the Dn4000 profile of galaxies hosting slow-strong bars, which show that they have the youngest stellar populations within the arms of the bar (although $<3\sigma$). This suggests that slow-strong bars fuel the most star-formation within the barred region of their SF hosts. 
    \item {\bf Star forming galaxies hosting fast-strong bars} (purple lines in Figure \ref{fig:sf_combo}), have profiles with the strongest dips in the bar-arms for EW[\ha], \hb, and \hd and a corresponding largest peak in Dn4000. The difference {in EW[\ha]} between fast-strong {(FSG)} bars and slow-strong {(SSG)} bars is {significant} throughout the bar-arms ({as indicated by the red triangles in the sub-panel of the upper leftmost panel in Figure \ref{fig:sf_combo} which show the difference between these two profiles)}, demonstrating that star-forming galaxies with fast-strong bars have more highly suppressed star-formation in their bar region.

\item {\bf Star-forming galaxies hosting weak bars} (both fast and slow; pink and blue lines in left panels of Figure \ref{fig:sf_combo}) have profiles that fall in between those of strong bars of different speeds.  In their bar region, they have higher sSFRs and younger stars than fast-strong bars but lower sSFRs and older stars than slow-strong bars. It is also notable that the difference between the profiles in galaxies hosting fast-weak and slow-weak bars are smaller than between fast-strong and slow-strong bars. The EW[H$\alpha$] profiles of fast-weak bars peak at the centre, decline within the arms of the bar, and peak {around} the bar-end as well; in contrast, the profiles of slow-weak bars have low values of EW[\ha] that rise almost uniformly towards the bar-end. While fast-weak bars may still trigger ongoing star formation at the centre and {around the} bar-ends in their star-forming hosts, there are no traces of long-lived impacts on their hosts that would have been distinguished in the profiles of post-starburst indicators (\hb and \hd) and Dn4000. Slow-weak bars exhibit an almost monotonic increase in SF with radius which suggests that they may not impact star-formation in their hosts on any timescale (as also noted in \citealt{geronthesis}). 

\end{itemize}

\subsubsection{Quenching Barred Galaxies}
\label{sec:quench_combo}

We also examine the radial trends in star-formation indicators for galaxies with the different bar types 
in quenching host galaxies. These are shown in the right column of Figure \ref{fig:sf_combo}. 
It is immediately notable how different these trends are from the star-forming barred galaxies. 
The profiles show that, at all radii, quenching barred galaxies have very low values of sSFR, \hb, and \hd at the centre. The SF activity then rises towards the outskirts of the galaxies, to be relatively comparable to (but still lower than) the levels found beyond the bar region in SF hosts. Similarly, at all radii, quenching barred galaxies have older stellar populations (higher Dn4000), which is particularly notable in the bar region. These broad trends in quenching barred galaxies were previously noted by \citealt{geronthesis} for the EW[\ha] and Dn4000 profiles. We probe the combined impact of bar strength and speed, finding it is observed most notably in the differences in the star-formation activity at the bar-ends as: 
\begin{itemize}
    \item {\bf quenching galaxies hosting slow bars (both strong and weak)} have low values of EW[\ha], \hb, \hd, and high values of Dn4000 at the bar ends.
    \item {\bf quenching galaxies hosting fast bars (both strong and weak)} have higher values of EW[\ha] and a steep gradient in \hd at the bar-end.
\end{itemize}
These observations suggest that active star formation is still ongoing around the bar end of quenching galaxies hosting fast bars whereas those hosting slow bars (both strong and weak) have been quenching more completely. These trends suggest that {\bf bar speed determines the extent of ongoing star-formation that occurs at the bar-ends}, with galaxies hosting slow bars more able to quickly deplete gas near the bar ends than those hosting fast bars.

In galaxies with fast bars, an additional impact of bar strength is observed - while the EW[\ha] profiles of fast bars show active ongoing star formation at the bar ends, this enhancement is especially large for fast-weak bars (pink lines). Within the bar region, quenching galaxies with fast-weak bars do not show any ongoing star-formation but have slightly higher levels of \hb and \hd (more recent star-bursts) and lower values of Dn4000 (young stellar populations) relative to other bar types in quenching hosts. Hence, we suggest quenching galaxies hosting fast-weak bars have only recently completed star-formation within the barred region and continue to have ongoing star-formation at the bar-ends. {\bf Galaxies with fast-weak bars therefore have the least difference in SF activity at the bar-ends between SF and quenching hosts than other bar types, suggesting their impact on their hosts is recent.}

\subsection{The Quenching Tracks of Different Types of Bars}
We have shown how the radial profiles of star-formation tracers provide insight into the spatial distribution of star-formation across the barred region at different timescales. {The \ha~and Dn4000 radial profiles presented in \cite{geronthesis} show that SF galaxies with slow-strong bars have the most impacted SF; we find further qualitative evidence for the enhanced impact of slow-strong bars on intermediate look-back timescales (through the \hd~and \hb~ tracers) and in evolved hosts (through the profiles of quenching galaxies, which show that those with slow-strong bars are more quenched than those with fast bars).} 

{In addition}, by plotting the median values of EW[$\ha$], \hb, and \hd as a function of the median values of Dn4000 in radial bins (see Figure \ref{fig:quenching_tracks}), we can gain more insight into the host galaxy evolution by {tracing how the different SF tracers vary over both time (i.e. in stellar population age/Dn4000) and space}. We refer to these 2D profiles as ``quenching tracks" since they show the progression of galaxies from young and actively SF to old and passive.

To construct these tracks, we use radial bins of size 0.3 \rb: twice the bin size of the 1D radial profiles; additionally, we only show four representative points of interest on the 2D profiles: (1) the {central galaxy region ($R \sim 0.15 \rb $), (2) the bar arms ($R \sim 0.75 \rb $), (3) the region just beyond the bar-end ($R \sim 1.35 \rb $) and (4) the galaxy outskirts ($R \sim 2-3 \rb $)}. Points are colored by radius (from {dark blue to light pink}; see Figure \ref{fig:quenching_tracks} legend) and connected by dotted/dashed lines. 

We do not have data for all barred galaxies to 3.0 \rb (light pink) -- bars that are fast and/or strong are longer relative to the galaxy's overall $R_e$, while MaNGA data ends at 1.5 or 2.5$R_e$. This has the result that galaxies with longer bars have MaNGA data that may tend to end shorter relative to the bar (around 2.0 \rb); therefore, the edges of the 2D profiles of different galaxies may vary slightly in shade. 

\subsubsection{Bar Impact on Central SFH}
As was previously evident in Figure \ref{fig:sf_combo}, we note that SF is enhanced at the centres of barred SF galaxies and almost absent in barred quenching galaxies. We additionally see {in Figure \ref{fig:quenching_tracks} (upper row)} that ongoing star-formation at the centre is enhanced in {SF} galaxies hosting strong bars (both slow and fast; {blue-filled circles and triangles respectively show high central EW[\ha]}), which have more recent/recently completed SF, and younger stellar populations than the centers of galaxies with weak bars {(the blue-filled circles and triangles respectively in the upper right panels Fig. \ref{fig:quenching_tracks} are not as high}). We also observe that, in SF strongly barred galaxies, those with slow-strong bars have slightly higher EW[\ha] than fast-strong counterparts. 
In quenching barred galaxies, we observe low sSFRs and old stellar populations at the centres {(see Figure \ref{fig:quenching_tracks}, upper row, unfilled symbols)}, although those hosting fast-weak bars {(rightmost column; blue-unfilled triangles)} show signs of more recently completed star-formation (higher \hb) and younger stellar populations (lower Dn4000).

\subsubsection{SFH within the bar-arms and around the bar-ends}
We observe that the tracks of SF galaxies hosting all bars types (except fast-weak) show an increase in Dn4000 and decrease in EW[\ha] towards the bar-arms, followed by an increase in EW[\ha], \hb, and \hd as well as a decrease in Dn4000 {just beyond} the bar-ends; this corresponds to the suppression of star-formation in the arms and triggered star-formation {around} the bar-ends that we also observed in the 1D radial profiles described in Section \ref{sec:rad_prof}. However, we observe notable differences in SF activity for strong bars with different bar speeds:
\begin{itemize}
    \item \textbf{Slow-strong bars in SF hosts} {(leftmost column, filled circles)} have the most enhanced and most recent SF {near} the bar-ends ({green markers, outlined in black}), as well as the youngest stellar populations. 
    \item \textbf{SF galaxies with fast-strong bars} ({middle-left column, filled triangles}) show the most decline in SF activity and increase in stellar age along the bar-arms ({teal markers}). Decreases in \hb and \hd along the arms are only observed for these bars in SF hosts, \textit{indicating that this suppression in SF is long-lived}. 
    \item \textbf{SF galaxies with weak bars (both fast and slow;} {middle-right and rightmost columns, filled circles and triangles respectively}) do not show a notable suppression within the bar-arms ({teal markers}); their tracks show a steady increase/only minimal decrease in star-formation from the centre ({blue markers}) to the bar-ends ({green markers, outlined in black}).
\end{itemize}   

All barred quenching galaxies show moderate increases in SF along the arms and towards the bar-ends ({unfilled circles/triangles in Figure \ref{fig:quenching_tracks}}). We see interesting differences in the extent of ongoing star-formation {just beyond} the bar-ends in quenching hosts relative to those in SF hosts for different bar types.
\begin{itemize}
    \item The ends of \textbf{slow-strong bars in SF and quenching galaxies} ({left column; green-filled and unfilled circles respectively}) generally occupy opposite extremes of the planes (upper left and lower right respectively). This suggests that \textbf{slow-strong bars have a significant impact on SF properties}, driving the tracks of SF host galaxies to the upper left corner (high SF activity, young stellar populations) and those of quenching host galaxies to the lower right corner (passive SF, old stellar populations) of these diagnostic planes.
    \item The extent of ongoing SF activity in \textbf{quenching galaxies with fast bars (both strong and weak}; shown in the {middle-right and right-most columns of Figure \ref{fig:quenching_tracks} with unfilled triangles}) are much closer in SF activity to that of these bars in SF hosts, \textit{indicating that a substantial amount of recent/ongoing SF is occurring in quenching galaxies with fast bars.}

\end{itemize}

\subsubsection{SFH in the outskirts of barred galaxies}

\begin{figure*}
    \centering
    \includegraphics[width=\textwidth]{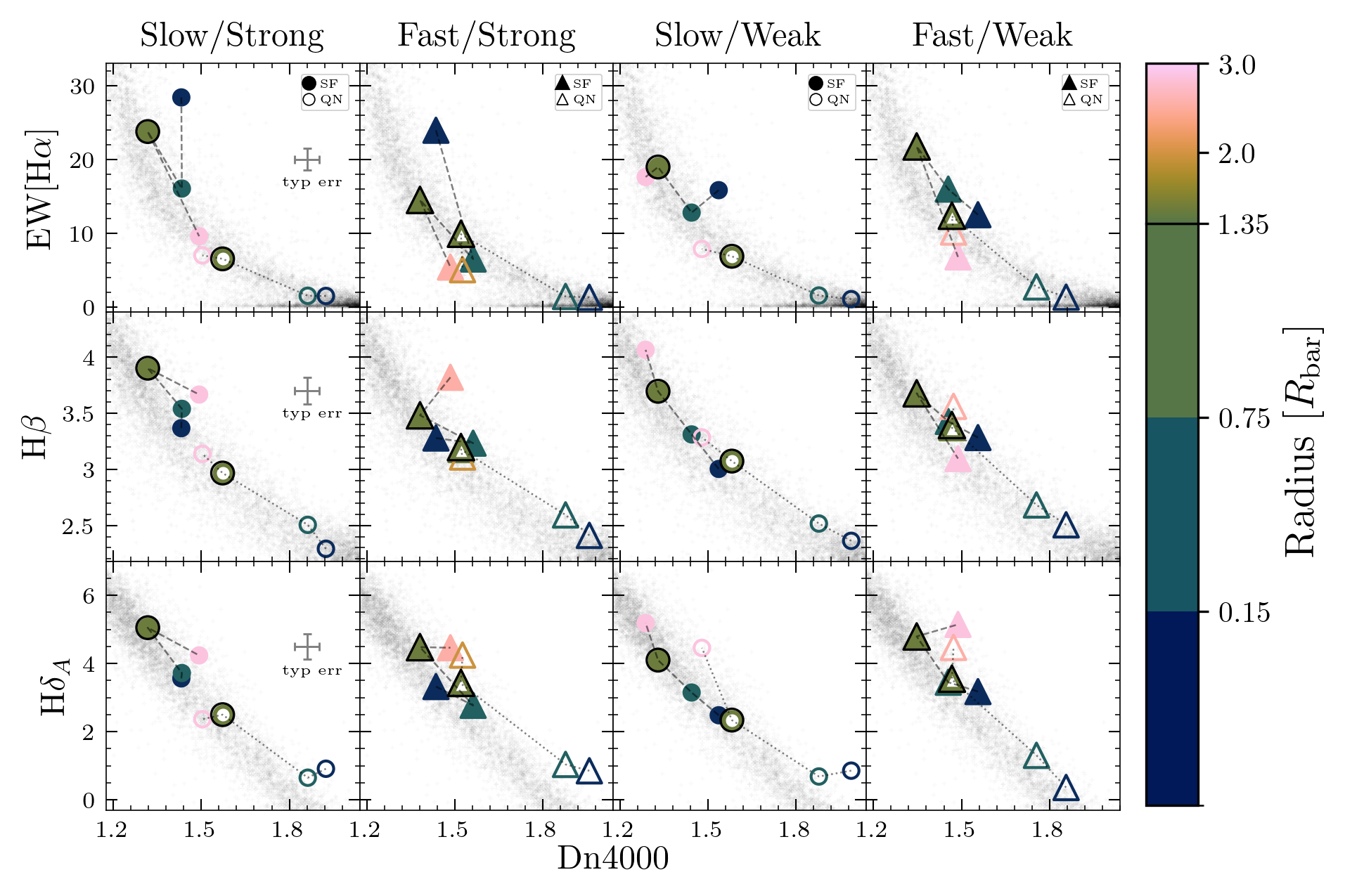}
    
    \caption{Median values ({circular/triangular colored markers}) of \ha, \hb, and \hd~ against those of Dn4000 in bins of size 0.3\rb (hereafter referred to as quenching tracks) for all eight subsets of the TW sample. Markers are colored by radius, with {dark} blue representing the galaxy centre and {light orange/pink} the galaxy outskirts. The bar-end region ({out to $\sim 1.35\rb$}) {corresponds to a green color and is enlarged and outlined in black for emphasis}. Each panel shows two quenching tracks for each bar group specified by bar strength and speed: the tracks for bars in SF and quenching host galaxies, with the corresponding median values shown in filled and unfilled markers respectively. Panels are arranged left-to-right by bar strength, with strong bars presented in the left columns and weak bars in the right. Bar speed is indicated by the marker shape, with fast bars shown in triangles and slow bars in circles. {Light gray background points show the 2D distribution of measurements of each star-formation tracer ( EW[\ha], \hb, and \hd) within 1 $R_e$ against those of Dn4000 for all 10,0001 MaNGA galaxies  \citep[obtained from the DAPAll catalog;][]{Westfall2019}. Note that the background distribution of each star-formation tracer against Dn4000 is the same across panels in the same row.} We observe that slow and strong bars show the most efficient evolution within the barred region, as star-formation is dramatically enhanced in SF host galaxies and effectively suppressed in quenching hosts.}
    \label{fig:quenching_tracks}
\end{figure*}

We observe that the outskirts of barred SF galaxies (indicated by the {orange}/pink filled markers {in Figure \ref{fig:quenching_tracks}}), with the exception of galaxies with slow-weak bars, have low sSFRs (decrease in EW[\ha]) and old stellar populations (increase in Dn4000). In most cases, the levels of ongoing (top panels showing EW[\ha] {in Figure \ref{fig:quenching_tracks}}) and recently completed (middle and bottom panels showing \hb and \hd) star-formation in the outskirts of quenching barred galaxies are similar to those of SF galaxies. The implications of this are twofold: (i) the impact of bars on their hosts' evolution is found only in the radial range containing the bar, and (ii) barred galaxies are primarily quenched inside-out. We note that the outskirts of quenching host galaxies and those of SF hosts are the most different in galaxies with slow bars ({orange/pink circles in the leftmost and middle-left columns}). 

\subsubsection{Global vs Local Impact of a Galactic Bar}
It is an open question if a bar has a global or local impact on the galaxy in which it resides. \cite{geronthesis} explored this question for different bar-types in the TW sample by comparing the radial trends of SF along the barred region to regions perpendicular to the bar; {thereby showing the 1D local difference of the distribution of SF tracers}. 

\begin{figure*}
    \centering
    \includegraphics[width=\textwidth]{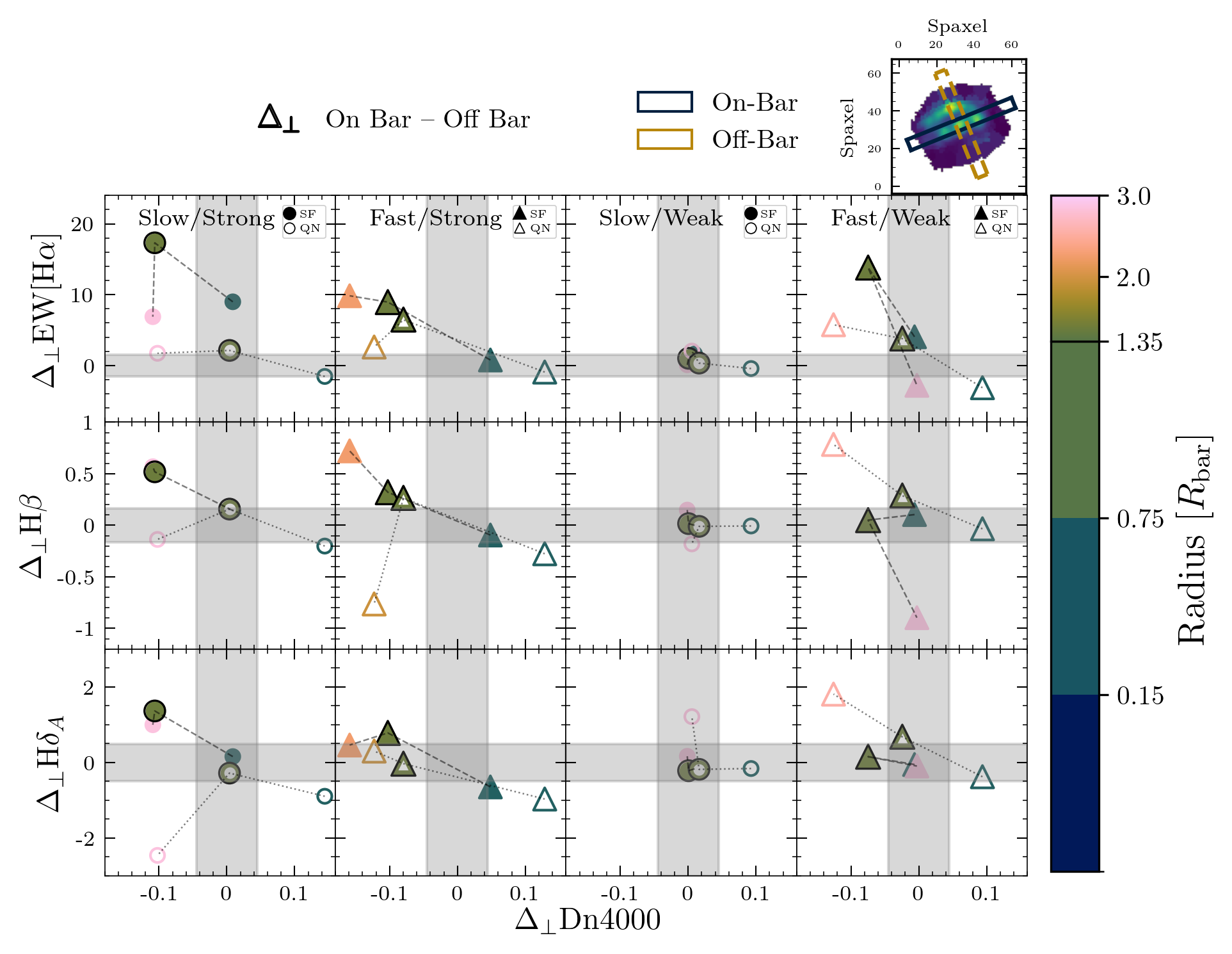}
    
    \caption{\label{fig:local_quenching_tracks}Residuals of quenching tracks in regions aligned with and perpendicular to the bar (i.e. the difference of the tracks in regions with the aperture placed on the bar and off the bar, as shown in the upper right inset figure) for all eight subsets of the TW sample. Panels and markers follow the same formatting as Figure \ref{fig:quenching_tracks}; however, the {average} values within the central bin ({$\sim$0.15\rb}) are omitted as the tracks overlap in the central region. Gray shaded regions represent the median Poisson errors on the radial distributions of each star-formation tracers over all eight bar groups, applied to a reference point around zero (so points within the gray shaded regions indicate no difference between the two alignments to 1$\sigma$). We see that, with the exception of slow-weak bars, the enhancement of star-formation at the ends of all bars in SF hosts is a local impact, but is especially notable in slow-strong bars.}
\end{figure*}

{In this work, we probe the 2D local variation of SF tracers by generating} residual median quenching tracks for each of our sub-populations of barred galaxies, shown in Figure \ref{fig:local_quenching_tracks}. Comparing these residual tracks allows us to further infer whether bars have localized impacts on galaxy evolution {(by analyzing the SF trends in both space and time/age)}. The procedure is as follows: 
\begin{itemize}
\item Following \cite{geronthesis}, we rotate our bar aperture to span the region perpendicular to the bar position angle (hereafter referred to as ``off-bar''; see upper right inset of Figure \ref{fig:local_quenching_tracks}). 
\item We create median radial profiles aligned with the bar position angle (hereafter referred to as ``on-bar'') and the ``off-bar'' alignment, with radial bins of size 0.3 \rb (the same as our other 2D quenching tracks and twice that of the 1D radial profiles).  \item We take the difference of the on-bar and off-bar radial profiles of each tracer to generate the residual profiles in Figure \ref{fig:local_quenching_tracks} - note that as the centers of both the ``on-bar'' and ``off-bar'' alignments overlap, we omit the central radial bin from the residual tracks and only show three representative regions of interest: 1) the middle of the bar arms (\rb $\sim$ 0.5\rb), (2) the bar-end (\rb $\sim$ 1.0 \rb) and (3) the galaxy outskirts (\rb $\sim$ 2.5-3\rb). 
\item We provide grey "error regions" in each sub-plot in Figure \ref{fig:local_quenching_tracks} - where the measurements are consistent with being identical to 1$\sigma$. 
\end{itemize}

We observe {similar trends to those of \cite{geronthesis}:} in barred SF galaxies, excesses in star-formation activity local to the bar alignment occur only at the bar-end or the galaxy outskirts. These excesses in star-formation vary significantly with bar-strength and bar-speed. {We add that some of these variations with bar-strength and bar-speed are observable on intermediate look-back timescales (through the \hb~and \hd~tracers).}
\begin{itemize}
    \item \textbf{SF galaxies with slow-strong bars} have bar-ends {(green-filled circles in leftmost column of Figure \ref{fig:local_quenching_tracks}, outlined in black)} that are consistently located in the upper left corner, revealing that these galaxies have the most enhanced ongoing and recent star-formation (traced by EW[\ha], \hb, and \hd) and young stellar populations (traced by Dn4000) at the bar-end relative to corresponding ``off-bar''/perpendicular regions in the same galaxy. \textit{This indicates that the enhancement in SF at their bar-ends is a local effect and is significantly different (to $1 \sigma$) from regions off the bar}, as also observed by \cite{geronthesis}.
    \item \textbf{SF galaxies with fast bars (both strong and weak)} show excess in ongoing star-formation (higher EW[\ha]) and younger stellar population at the bar-ends relative to off-bar regions, although this excess is greater for fast-strong bars {(see green-filled triangles in upper row, middle right and rightmost columns in Figure \ref{fig:local_quenching_tracks}, outlined in black)}. However, this local impact at the bar-ends is not significantly different over longer timescales (in the planes of \hb and \hd against Dn4000; {middle and bottom rows of Figure \ref{fig:local_quenching_tracks} respectively}).
    \item \textbf{SF galaxies hosting slow-weak bars} have residual tracks that do not show differences in the star-formation activity of on-bar and off-bar regions (within 1$\sigma$).
    \item {\textbf{Quenching Galaxies with fast-strong bars} have enhanced ongoing SF at the bar-ends that is local to the bar region: the residual ongoing SFR at the bar-end (green-unfilled triangle in the upper row, middle-left column of Figure \ref{fig:local_quenching_tracks}, outlined in black) shows a $\,>$$1\sigma$ difference between the on-bar and off-bar region. We only see marginally local effect of more recently completed star-formation (through slightly enhanced \hb) at these bar-ends (black-unfilled triangle in middle row, middle column of Figure \ref{fig:local_quenching_tracks}).}
\end{itemize}
While many barred galaxies show significant differences in SF activity at the galaxy outskirts ({orange/pink circles and triangles}), these differences vary sporadically between excesses and suppression of SF and do not show consistent trends with bar-type; it is therefore unclear what drives these differences in SF activity between the outermost regions of the galaxy aligned with the bar and those perpendicular. The outskirts of even barred galaxies may be asymmetric for a variety of reasons not related to the bar itself. 
\subsection{Measures of Correlation: Interaction with Other Physical Properties}
\label{sec:correlations}

To further probe the extent of bar-driven impacts on the SFH of galaxies, in this section, we analyse correlations between the shape of the radial profiles of star-formation indicators in barred galaxies with other bar and galaxy properties. We define four representative quantities to capture the shape of star-formation radial profiles: (1) the value at the centre ({average within the central 0.01 \rb}), (2) the change between the central value to the middle of the bar ($\Delta_{\mathrm{Arms-Centre}}$, {with the middle of the bar corresponding to the average value between 0.4-0.6\rb}), (3) the change between the values at the bar-end and mid-bar (or ``arms'' $\Delta_{\mathrm{End-Arms}}$, {with the bar-end corresponding to the average value between 0.9-1.1\rb}), and (4) the change between the values in surrounding outskirts beyond the bar-end to those at the ends ($\Delta_{\mathrm{Beyond-End}}$, {with the outskirts beyond the bar-end corresponding to the average value between 1.3-1.7\rb}). We evaluate correlations between these measurements and various physical properties of the bar and host galaxy. These are: (1) bar-speed (as measured by the continuous variable $\mathcal{R}$, recall galaxies with $\mathcal{R}>1.4$ are ``slow'' and those with smaller $\mathcal{R}$ are ``fast''), (2) bar-strength (quantified by defining the difference of the GZ DESI strong bar and weak bar vote fractions as $p_{\mathrm{strength}}=p_{\mathrm{strong}}-p_{\mathrm{weak}}$), (3) specific star formation rate (sSFR) and (4) stellar mass, both obtained from MaNGA data via Pipe3D \citep{Sanchez2016}, (5) galaxy environment, as measured by the tidal strength to the first nearest neighbor (from \citealt{Argudo2015}), and (6) HI deficiency (from HI-MaNGA; \citealt{Masters2019}).

We examine all pairwise correlations between bar strength, bar speed, other galaxy properties and our measures of the shape of the radial profile of star-formation indicators across the barred region for our sample. {These are illustrated in Appendix Figure \ref{fig:correlations} which shows all the correlation coefficients, samples sizes and $p$-values with relationships having significant correlations indicated by the green hi-lighted boxes.} 
 
The significant correlations that we observe ({i.e. those for which the null hypothesis can be rejected, or which have $p$-values, i.e. probability of the null hypothesis,} $p < 0.03$) {include}:
\begin{itemize}
    \item The central value of all SF tracers correlate (as would be expected) with overall sSFR of the galaxies, and are inversely correlated with the galaxy mass (Dn4000, being larger for older stellar populations, is inverted - i.e. increased with stellar mass, and decreases with increasing sSFR). 
    \item {The magnitude of the change in recent SF (\hb and \hd) in the outskirts is observed to correlate with sSFR and stellar mass, such that larger changes in recent SF towards the bar-ends and outskirts (i.e. $\Delta$\hd$_{\rm End - Arms}$ etc.) being associated with higher masses and lower sSFRs. As seen in Figure \ref{fig:sf_combo}, we note that the gradients of the profiles of recent SF tracers for quiescent galaxies are larger than in star-forming hosts which is reflected in these correlations.}

    \item  $\Delta$ EW[\ha]$_{\rm Arms - Centre}$ is found to be more negative (and $\Delta$Dn4000$_{\rm Arms - Centre}$ more positive in this region) when bars are stronger -- revealing how strong bars suppress SF along their length. {Additionally $\Delta$Dn4000$_{\rm Arms - Centre}$ is positively correlated with sSFR - SF barred galaxies have low Dn4000 at all radii, while there is a notable increase in the amplitude of the negative gradient of this tracer in quenching barred galaxies.}
    \item The $\Delta_{\rm End - Arms}$ of tracers show a variety of correlations {mostly} with bar properties; bars clearly having a significant impact on the SF of galaxies in the radial region from their mid point to their end. For {ongoing} SF (EW[\ha]), the {only significant} correlation is with bar speed, $\mathcal{R}$ suggesting a bigger difference between {current} SF in the arms and the ends of faster bars $1<\mathcal{R}<1.4$ relative to slower ones $\mathcal{R}>1.4$ -- showing that fast bars have less SF in the bar. {The change in recently completed SF ($\Delta$  $\rm{H\beta_{End - Arms}}$) shows similar correlations, with faster bars (lower $\mathcal{R}$) and stronger bars having a bigger difference between recent SF in the arms and the bar-ends, highlighting that fast-strong bars in particular have suppressed SF in the arms.} On the other hand, the change in Dn4000  (which is largely negative in this region, hence a positive correlation means a smaller change {with larger values of the correlating variable}) shows correlations with bar speed and strength -  {revealing a trend towards steeper outer declines - i.e. more younger stellar populations for slow-weak bars, and in lower-mass galaxies.}

\end{itemize}
We do not observe any significant correlations between the turnovers in star-formation and our measure of environment ($Q$) {and only one correlation between HI deficiency and stellar age beyond the bar-end (older stellar populations beyond the bar associated with more HI deficient hosts)}. The overall shape of barred galaxy SFH profiles is thus more dependent on their internal properties (stellar mass, and bar properties) than environment or gas content. 

Overall, the correlations provide quantitative evidence that star-formation in the arms, or bar region of a barred galaxy depends largely on bar-speed and strength, such that star-formation is especially suppressed in the arms of bars that are both fast and strong, but can still occur throughout the barred region in slow-strong bars.

\subsection{Connections to \ha~Morphology}\label{sec:hamorph}
Previous authors \citep[e.g.][]{McKelvie2020} have suggested star-formation rings are linked to bars. The presence of rings could indeed explain our observations of enhanced SF just beyond the bar-ends, particularly for cases where the enhancement in SF is not local to the bar-end (galaxies that have enhanced SF in the off-bar alignment as well as the bar-end, such as those with slow-weak or fast-weak bars; see e.g. Figure \ref{fig:local_quenching_tracks}). 

\begin{figure*}
    \centering
    \begin{tabular}{c c}
        \begin{subfigure}{0.53\textwidth}
            \centering
            \includegraphics[width=\textwidth]{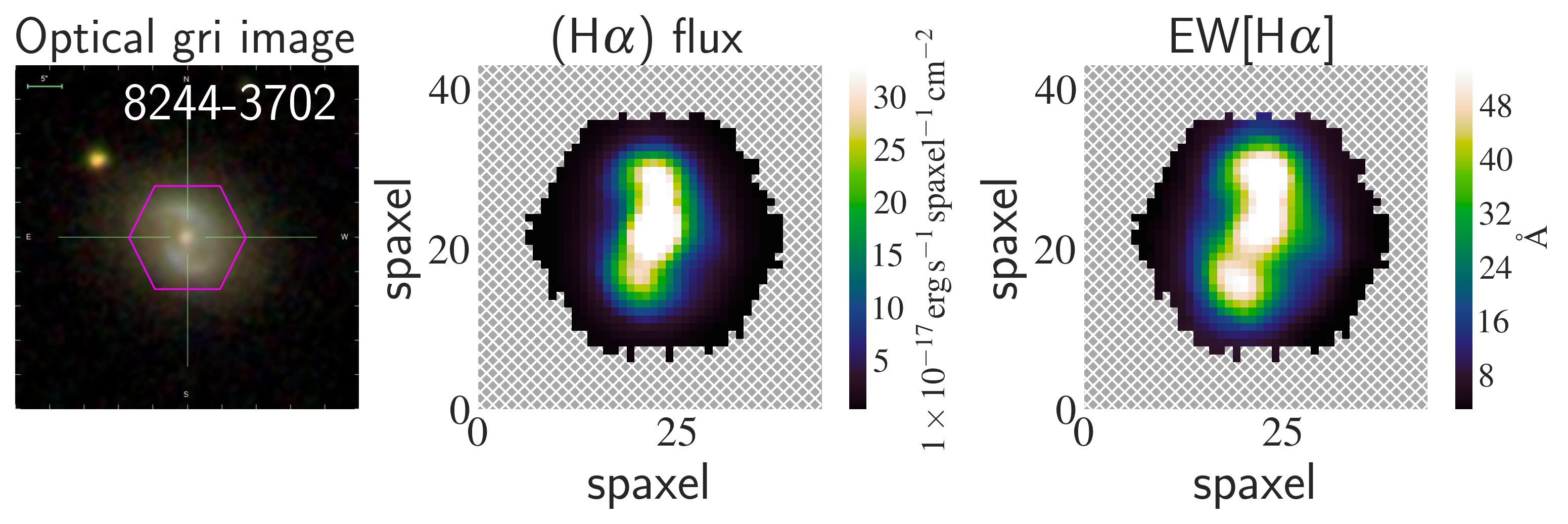}
            \caption{H$\alpha$ along bar}
        \end{subfigure} 
        & 
        \begin{subfigure}{0.53\textwidth}
            \centering
            \includegraphics[width=\textwidth]{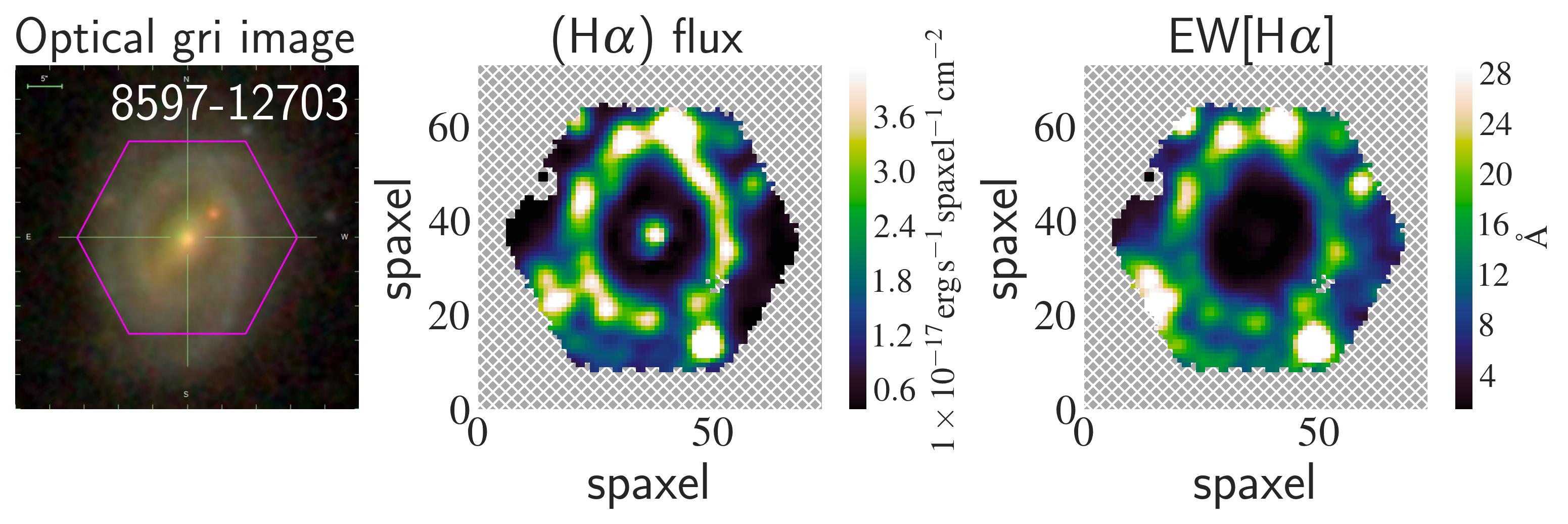}
            \caption{H$\alpha$ present in a prominent ring}
        \end{subfigure} \\
        
        \begin{subfigure}{0.53\textwidth}
            \centering
            \includegraphics[width=\textwidth]{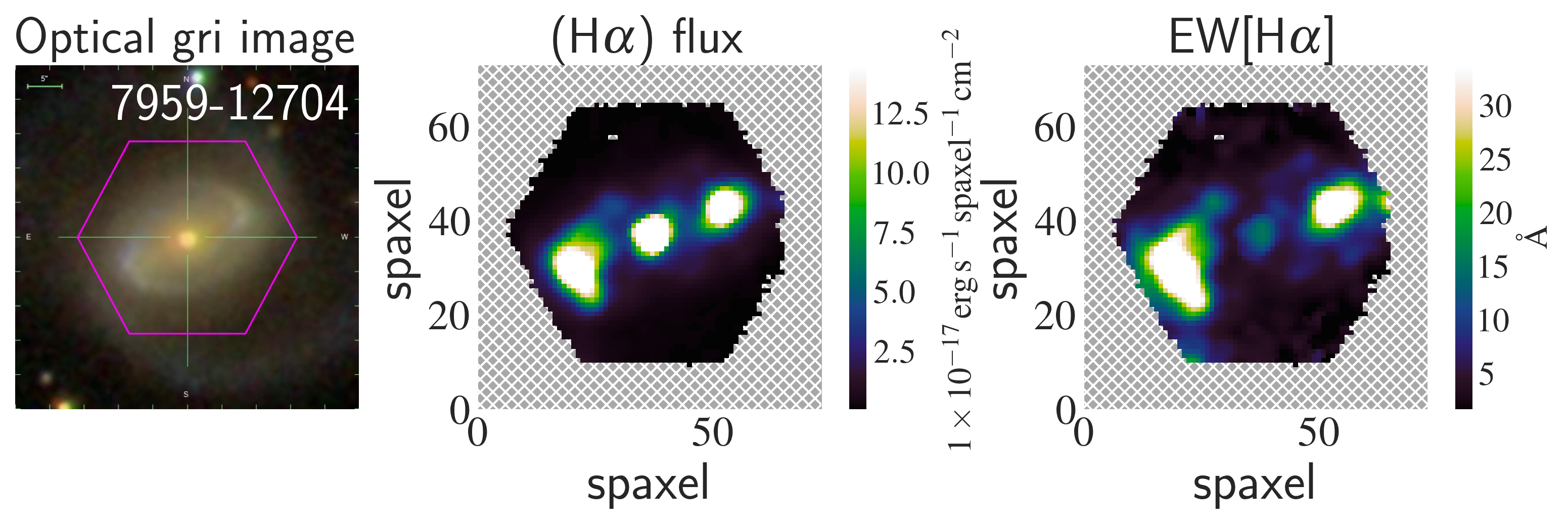}
            \caption{H$\alpha$ concentrated at bar-ends in addition to centre.}
        \end{subfigure} 
        &  
        \begin{subfigure}{0.53\textwidth}
            \centering
            \includegraphics[width=\textwidth]{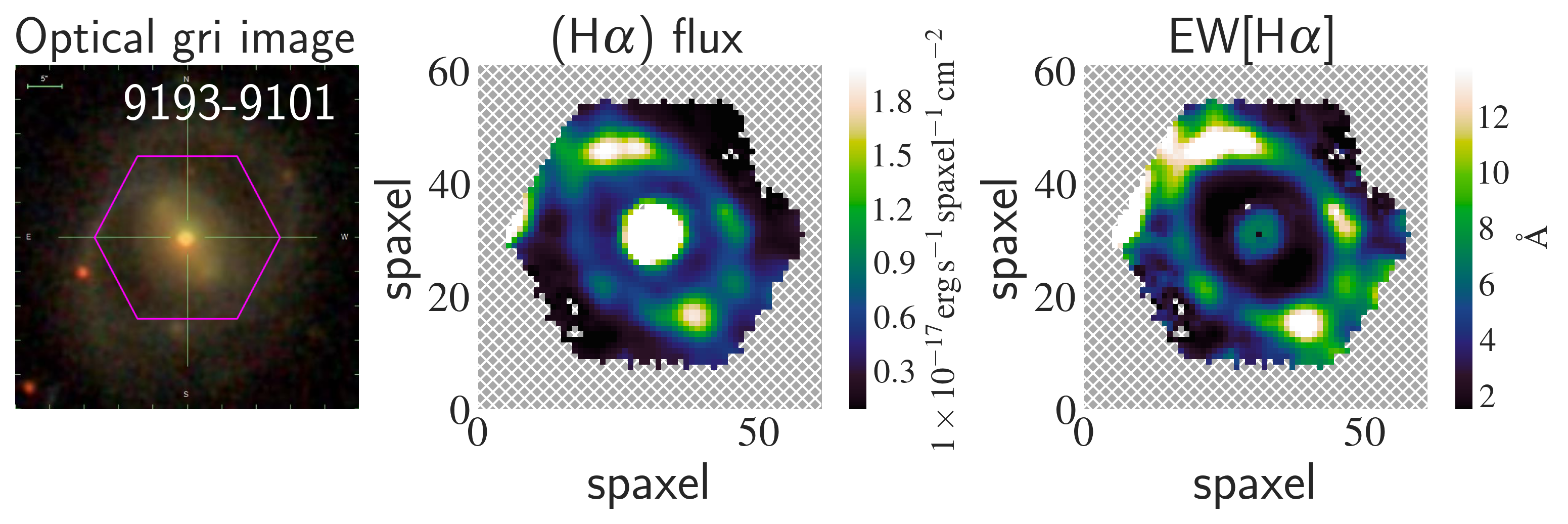}
            \caption{H$\alpha$ present in a ring, with concentrated nodes at the bar-ends}
        \end{subfigure} \\

        \begin{subfigure}{0.53\textwidth}
            \centering
            \includegraphics[width=\textwidth]{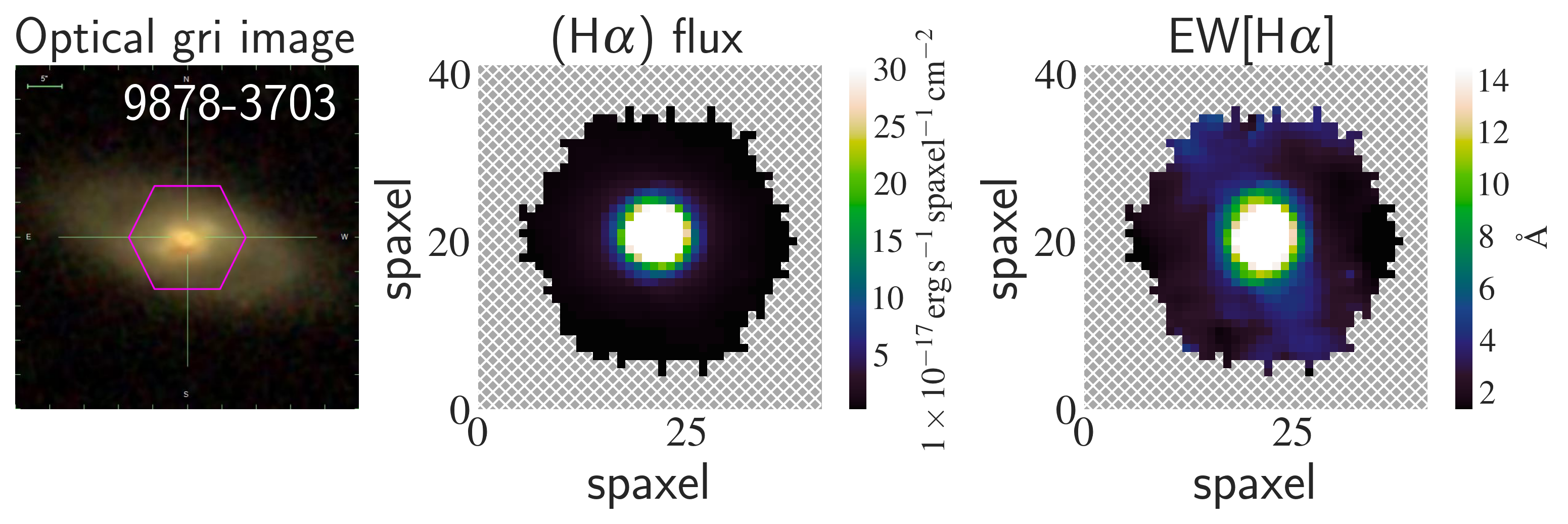}
            \caption{H$\alpha$ predominantly at centre}
        \end{subfigure} 
        & 
        \begin{subfigure}{0.53\textwidth}
            \centering
            \includegraphics[width=\textwidth]{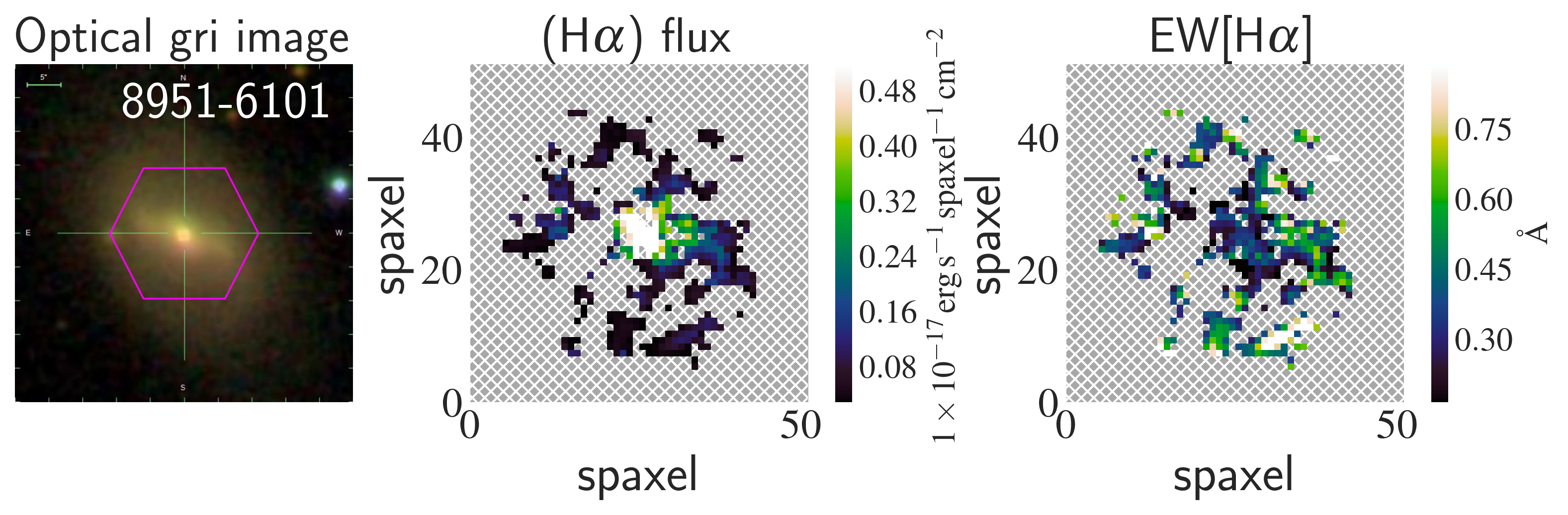}
            \caption{No significant H$\alpha$ }
        \end{subfigure} \\

        \begin{subfigure}{0.53\textwidth}
            \centering
            \includegraphics[width=\textwidth]{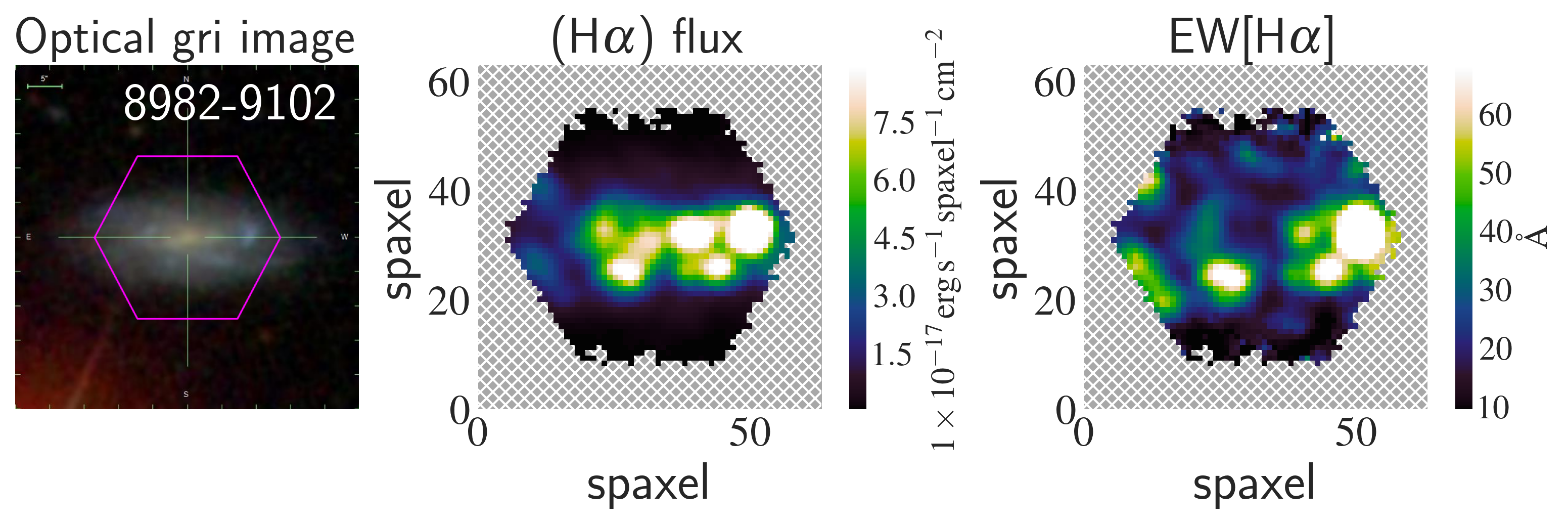}
            \caption{Unclassifiable}
        \end{subfigure} 
        &
        \begin{subfigure}{0.53\textwidth}
            \centering
            \includegraphics[width=\textwidth]{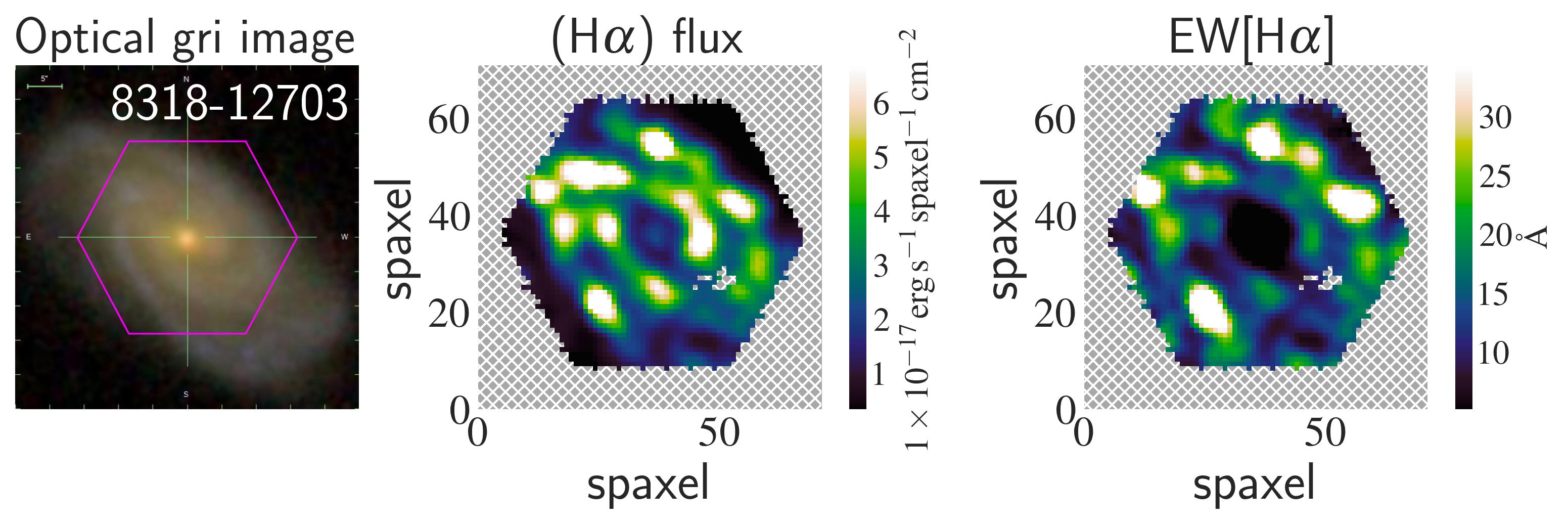}
            \caption{H$\alpha$ present mostly in spirals}
        \end{subfigure}
    \end{tabular}
    \caption{Examples of each \ha ~morphological category defined in \protect\cite{McKelvie2020} and two additional categories introduced in this work (\textit{In a ring and at bar-ends} and \textit{Spirals}; see panels \textit{(d)} and \textit{(h)} respectively) defined for galaxies within the TW sample. Panels show the optical image for an example galaxy (left sub-panel) along with the map of its Gaussian Emitted \ha ~flux (central sub-panel) and EW[\ha] (right sub-panel).}
    \label{fig:mckelvie_classifications}
\end{figure*}

\begin{figure}
    \includegraphics[width=0.5\textwidth, left]{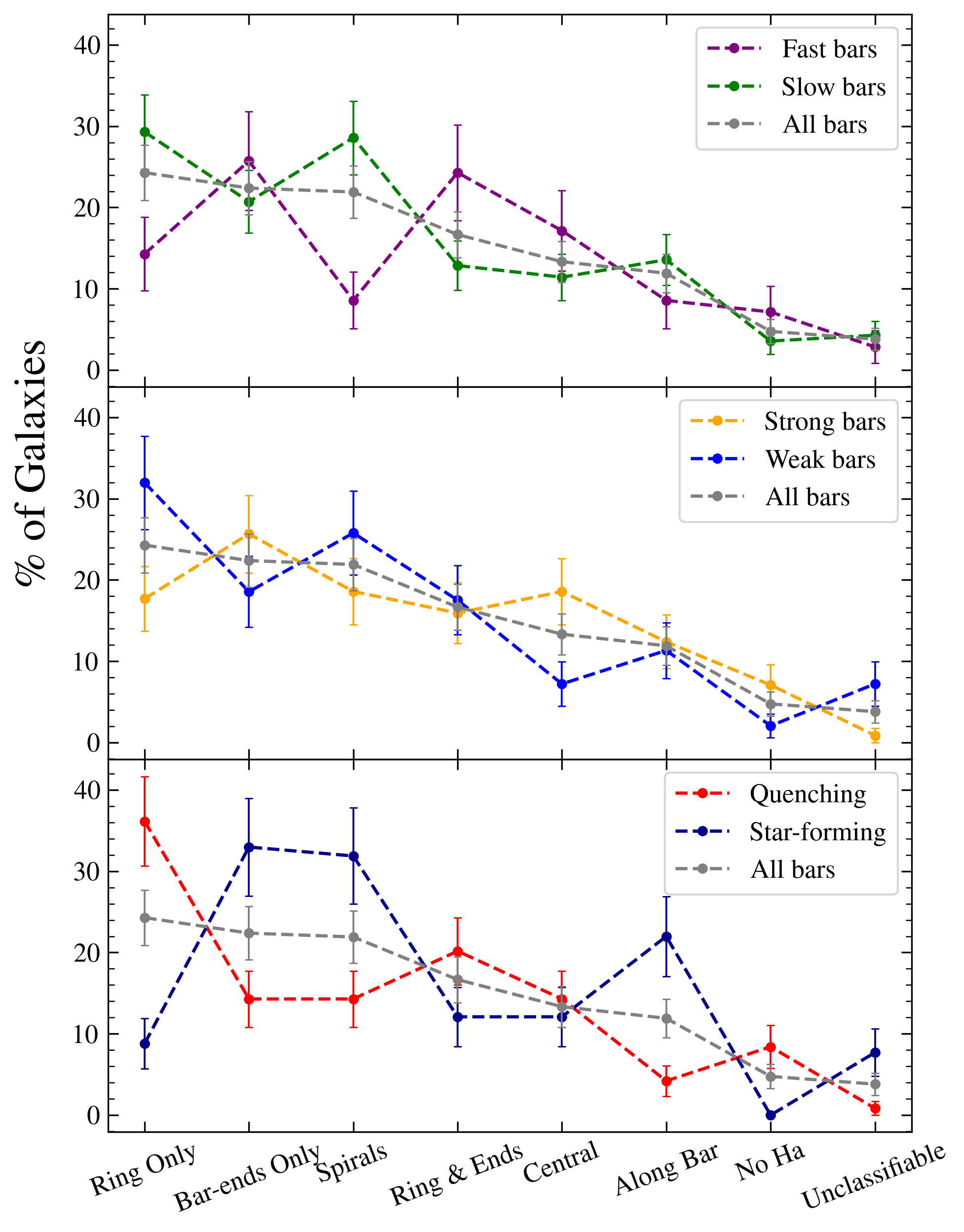}
    \caption{Classifications of \ha ~morphology (EW[\ha] and \ha ~flux) for the TW sample, shown for subsets split based on bar speed (fast vs slow), bar strength (strong vs weak), and SFRs (star-forming vs quenching), with properties of the entire sample shown for comparison in the gray dashed lines. Error bars indicate the standard Poisson errors on the classifications. Symmetric rings and spirals are more common in slow bars relative to fast bars. Central \ha ~and concentrations in \ha ~nodes at the bar-ends are more likely to be found in strongly barred galaxies. \label{fig:mckelvie_classifications_results}}
\end{figure}

\citet{McKelvie2020} identified various \ha ~morphologies in barred galaxies with MaNGA data. We extend this classification to our entire sample, with two of the authors (PM, KLM) independently classifying all galaxies, and revisiting those where we disagreed until consensus was obtained. Examples of these classifications are shown in Figure \ref{fig:mckelvie_classifications}. We consider whether any of the subtypes of barred galaxies (strong or weak; fast or slow; SF or quenching) show  any trends in these types of \ha ~morphology. Our results are visualized in Figure \ref{fig:mckelvie_classifications_results}. Overall, we find that in the entire TW sample {(see gray lines in Figure \ref{fig:mckelvie_classifications_results}; the gray line in each panel is the same and represents the morphology of the entire TW population)}: 11.9\% of barred galaxies show \ha ~along the bar, 13\% are dominated by central \ha, 41\% of barred galaxies have a \ha ~ring (24.3\% a symmetric ring, and 16.7\% a ring with \ha ~nodes); \ha ~nodes at the end of a bar are found in 39\% of barred galaxies (16.7\% with a ring, and 22.3\% just have nodes), 21.9\% of barred galaxies show a \ha ~spiral, and about 8.6\% total have either no \ha ~or unclassifiable/irregular \ha.  

Some of the features (e.g. \ha ~along the bar, no \ha, or unclassifiable \ha ~morphology) show no significant differences between different types of bars, although these are different between SF and quenching bar host galaxies ({see red and blue lines respectively in lower row of Figure \ref{fig:mckelvie_classifications_results})}, as it is more likely to see \ha ~in SF galaxies: it is also much more likely to see \ha ~along the bar in barred galaxies with active SF. Features that are found to be notably different in different sub-samples are: 

\begin{itemize}
 \item Smooth \ha ~rings and spirals are more likely to be found in galaxies with slow bars than fast bars {(green and purple lines respectively in Figure \ref{fig:mckelvie_classifications_results}, upper row)}, which is consistent with recent evidence implying that slow bars are more commonly associated with nuclear structures \citep{Lee2025}. Galaxies with fast bars have more \ha ~rings with nodes at the end (noting that, averaged over both kinds of rings, the two types of bars are equally likely to have some kind of ring; about 40\% have rings in both cases). 
 \item Strong bars {(orange lines in Figure \ref{fig:mckelvie_classifications_results}, middle row)} are found in galaxies with fewer \ha ~rings and more \ha ~in very bright central regions (relative to the rest of the galaxy) and in bar-end nodes compared to weak bars {(blue lines in Figure \ref{fig:mckelvie_classifications_results}, middle row)}. 
 \item Quenching galaxies with bars {(red lines in Figure \ref{fig:mckelvie_classifications_results}, lower row)} are more likely to have \ha ~rings, and less likely to have \ha ~along the bar, at bar-end nodes, or in a spiral than SF galaxies with a bar {(blue lines in Figure \ref{fig:mckelvie_classifications_results}, lower row)}. 
\end{itemize}

Tying this together, it is clear that galaxies that have slow and/or strong bars and are still SF have the most rings and spirals in \ha. This is revealing enhanced SF, including SF along the bar, and nodes at the end. Barred ringed galaxies are observed to be more quenched \citep[lower gas fractions and redder colors][]{Fernandez2025}; hence, the \ha ~morphology of slow and/or strong bars provides further evidence of their heightened impact on their hosts' evolution. Due to the limited size of our sample, we cannot make statistical conclusions about the \ha ~morphology of samples split based on combinations of bar speed, bar strength, and host SFR. However, we note that, out of 26 slow-strong bars in SF hosts, 11 have SF concentrated at the bar-ends only, and 13 have \ha ~along the bar. These fractions are consistent with the enhancements in SF along the bar and local impact of the ends of slow-strong bars that we observe throughout this work. In quenching galaxies, rings of ongoing SF outside the bar, especially in weak bars, are common in the \ha ~morphology. These rings appear symmetric in galaxies with slow bars, while in those with fast bars, they also contain enhancements (or nodes) lined up with the bar ends, which is also consistent with the SF activity observed at the ends of fast bars in quenching galaxies. In future work, it would be interesting to consider how these \ha ~morphologies correlate with deviation from the typical star forming sequence of galaxies.

\section{Discussion} 
\label{sec:discussion}
We have shown that the distribution of radial profiles of different star formation indicators for galaxies hosting different types of bars changes qualitatively throughout the barred region and that turnovers in star-formation correlate significantly with bar strength and bar-speed in localized regions. Here, we discuss the physical impacts that different types of bars may have on their hosts.

\subsection{Radial Inflow and Starbursts at the Centre}
We clearly observe increased star-formation activity at the centres of star-forming barred galaxies.  While bar speed does not notably influence the extent of this central star-formation, bar strength has clearly observable impacts: stronger bars drive the largest enhancements in central star-formation in SF host galaxies. SF galaxies with fast-weak bars also show signs of increased ongoing star-formation at the centre (through higher EW[\ha]) but no long-lived traces of central starbursts, suggesting the impact of these bars are very recent. In already quenching barred galaxies, all bars -- regardless of type -- have low central sSFRs and evidence for both long-completed star-formation, and old stellar populations at their centres, implying that they have been centrally quenched (see Figure~\ref{fig:sf_combo}). 

Our findings support the already substantial evidence in the literature that bars are linked to the elevated SFRs at the centre of barred galaxies. Central SFRs of barred galaxies are typically found to be boosted \citep[see e.g.][]{Ellison2011,George2019a,  George2020, Wang2020, Newnham2020}. Similar trends are also seen in radial gradients of star-formation tracers in other IFU surveys: a high fraction of barred galaxies (88 $\pm$ 8 $\%$) in both the CALIFA \citep[Calar Alto Intergral Field Area;][]{Sanchez2016califa} and MaNGA surveys have radial gradients of EW[\ha], \hd, and Dn4000 that show central enhancements in the sSFRs that can reach up to an order of magnitude increase  \citep{Lin2017, Lin2020, geronthesis}.

The most plausible process by which bars influence central star-formation is the radial inflow of gas that occurs as the bar rotates through its host. This model is supported both by observations of gas flow \citep[see e.g.][]{Casasola2011, Combes2013} and simulations \citep{Athanassoula1992a,Villa-Vargas2010, Spinoso2017}.  As gas is funneled inwards towards the centre, this sparks central SF, and increases central mass concentrations (CMCs) for stronger bars \citep{Villa-Vargas2010,Athanassoula2013}. This relation between bar strength and radial inflow sparking central SF easily explains the trends seen in this work of larger central SF enhancements in galaxies with strong bars. {In addition, interactions between bars and the gaseous component of the disc show that bar growth is favored in gas-poor discs; as bar growth is positively correlated with bar strength, this implies that stronger bars are more likely to exist in quiescent galaxies (which are likely to have less gas; also see \citealt{Masters2012}), and may therefore explain why we see more strong bars among the quenching barred galaxies}.

There is also considerable literature that shows not all barred galaxies have enhanced central SF \citep[see e.g][]{Scaloni2024, Renu2025}. This is consistent with the star-formation histories observed for the quenching/quiescent population in this work. We demonstrate that globally SF barred galaxies show central starbursts while quenching barred galaxies have a lack of central SF \citep[also see ][]{geron2021, geronthesis}. This suggests that barred galaxies may generally follow an inside-out quenching process \citep[see e.g.][]{Tacchella2015,Lin2019}. 

\subsection{Turbulence, Shear, and Suppression of Star-formation in the Bar Arms}

Within the arms of the bar, we observe notable variations in the SF properties of the host galaxies which correlate with bar strength and bar speed. Barred SF galaxies with slow-strong bars have consistently higher levels of SF in the arms and younger stellar populations than other bar types in SF barred galaxies. Conversely, SF galaxies with fast bars (both strong and weak) have the most suppressed SF in their bar regions  (see Figure \ref{fig:sf_combo}; with fast-strong bars having the most suppressed SF). 

We observe that it is bar strength, even more than galaxy stellar mass, which has the strongest correlation with steep declines in ongoing SF (EW[\ha]) along the bar arms. Hence, we suggest that bar strength drives the suppression of SF in the arms of a bar while bar speed influences the extent of the suppression: strong bars show the most suppressed SF along the bar, particularly when these bars are also fast, whereas SF can still occur along the arms of slow bars, even if they are strong.   

Simulations have shown that bar-arms can be left as a region of lower SF activity in barred galaxies \citep{Tubbs1982,Sheth2000, Momose2010, Watanabe2011, Gavazzi2015,James2018}. The dominant mechanism is via increased velocity dispersion of gas molecules caused by shear motions driven by the the bar potential \citep{Khoperskov2018, Maeda2023, Kim2024}. Shocks (caused by the motion of leading edges of the bar relative to the gas in the rest of the disc) are predicted to drive angular momentum flux of the gaseous component in the disc, increasing turbulence and thereby maximizing shear along the arms \citep{Zurita2004, Kim2012a, Emsellem2015, Kim2024} which, in turn, inhibits cloud collapse for SF \citep{Elmegreen&Scalo2004, Sun2020}. There is even evidence for cloud collisions in some nearby barred galaxies and where the collisional velocities are high, SF is suppressed \citep{Fumiya2025, Kolcu2025}. The bar-shock suppression of SF is predicted to increase with increasing bar strength and length \citep{Khoperskov2018}. {Fast bars are relatively longer than slow bars with respect to the corotation radius (i.e. fast bars reach out to co-rotation whereas slow bars end shorter than this radius)}, and strong bars are longer than weak bars almost by definition. Hence, {fast-strong bars are the longest bars in the TW sample relative to the corotation radius;} the enhanced suppression in the arms we observe for galaxies with fast-strong bars can be explained by these bars {have} the highest shear due to their {increased length relative to corotation}. \citet{Athanassoula1992a} consider bar shocks in models with different "Lagrangian radii" (which \citet{Athanassoula1992} explain is roughly comparable to the co-rotation radius {even} for strongly barred galaxies). \citet{Athanassoula1992a} suggests bars which are comparable in length to the Lagrangian radii ({i.e. fast bars}) generate the strongest shocks, which explains our observations if we interpret the suppression of SF in fast bars as due to shocks from a bar reaching out to near its Lagrangian radius.

\subsection{Gas Cloud Collisions and Star Formation at the Bar-End}
We observe that all SF barred galaxies in our sample show ongoing/recently triggered SF at the bar-ends. Galaxies with fast-strong bars have low levels of SF at the bar-ends whereas those with slow-strong bars have the highest SF activity in this region, as observed by their radial profiles in Figure \ref{fig:sf_combo} and their locations (lower and upper left respectively) in the evolutionary planes presented in Figure \ref{fig:quenching_tracks}. We note that the enhanced SF at the ends of slow-strong bars is localised to the bar-ends (see Figure \ref{fig:local_quenching_tracks}), indicating that the SF is particularly enhanced at these bar-ends relative to a surrounding ring at this radius. The stellar populations and SF activity of weak bars fall in between those of strong bars and show little variation with bar speed. In quenching host galaxies, we observe that triggered SF at the bar-end persists for those hosting fast bars (both strong and weak), although the extent of SF activity is greater at the ends of fast-weak bars. Quenching galaxies hosting slow bars have little to no star-formation activity and older stellar populations, which suggests that slow bars may be so efficient at triggering SF at the bar-end in their hosts when they are still forming stars that these galaxies quench more completely.

The enhancement of SF at bar-ends has been widely explored \citep[e.g.][]{Renaud2015,Emsellem2015, Ansar2023}. The rotation of bars results in turbulence and shocks at bar-ends and edges; this turbulence is balanced by the shear induced in the ISM at the bar-end, which allows for the formation of dense clouds and increases orbital crowding \citep{Renaud2015,Emsellem2015, Ansar2023, Fumiya2025}. Consequently, {the number of} gas cloud-cloud collisions increase, suppressing SF in the bar-arms \citep[see e.g.][]{Fujimoto2020} and triggering star-formation activity at the bar ends \citep{Renaud2015}. Signs of SF enhancements at the bar ends have been observed in the \ha~morphology of massive barred galaxies, which show concentrations of \ha~at the bar-end \citep{Reynaud1998, Verley2007, Neumann2019,McKelvie2020}; these galaxies are more likely to host longer and/or stronger bars \citep[see e.g.][]{Nair2010bars,Masters2011, McKelvie2020, Mukundan2025}, which explains the increased occurrence of \ha~concentrated at bar-end nodes in galaxies with strong bars that we see in our sample (see Section \ref{sec:hamorph}). Our observations of local excesses in triggered SF at the ends of slow-strong bars in SF galaxies in the radial profiles and quenching tracks also provides further evidence that the extent of triggered SF at the bar-ends increases with bar strength. 

\subsection{The Bar of the Milky Way}
{It is remarkably challenging to place the bar of our own Milky Way (MW) into context with the fast-slow; strong-weak continuum of bars presented here. The MW bar is often described as weak, and assumed to be fast (i.e. assuming $\mathcal{R} = 1$), which suggests it should have limited impact on the SF in the MW, but as has been noted by many previous authors (e.g. \citealt{Merrifield2004,Bland-Hawthorn2016,GaiaCollaboration2023}) constraining either the strength, length or the kinematics of the Galactic bar based on observations of it from our vantage point provides significant challenges; there has been at least one suggestion of evidence that the MW bar could be long (i.e. most likely strong) and slow \citep{Hunt2018}. It is also challenging to constraint the global SF properties of our own Galaxy, although there is good evidence the MW is likely quenching \citep{Fielder2021}, and that that process began at a similar time as the MW bar formed \citep{Haywood2016}. Even though our vantage point within the MW makes it challenging to comment on how the picture we present ties into the ways that our own bar has impacted (or continues to) the SF history of our own galaxy, it is clear, even from a quick review of soem of the most recent published work in this area, that the bar of the Milky Way is connected to significant radial and/or non-axisymmetric flows \citep[most recently constrained using Gaia data][]{GaiaCollaboration2023}, and may be associated with over/underdensities in young stars \citep{Feng2026} and molecular gas \citep{Evans2026}, suggesting a clear impact on the Galactic SFH. }

\section{Conclusions}
\label{sec:conclusion}
Using the sample of 210 MaNGA galaxies with bars and bar pattern speeds derived by \cite{geron2023}, we generate radial profiles along the direction of the bar for four tracers of star formation in nearby galaxies, namely EW[H$\alpha$] for recent SF, H$\delta_A$ and H$\beta$ for recent post-starburst tracers, and Dn4000 to approximately trace stellar population age. \cite{geronthesis} explored the combined impact of bar strength and speed by generating radial profiles of two of these tracers (EW[\ha] and Dn4000) for subsets of SF galaxies with bars of different strengths and speeds. We continue this work by further probing the combined impact of bar strength and speed. We compare the radial profiles of both SF and quenching galaxies hosting combinations of strong/weak and fast/slow bars, including two additional tracers of SF on intermediate timescales to get more insight into the bar's impact across the host's history of SF evolution. Additionally, we trace the evolution of all eight subsets of barred galaxies by examining the co-variance of tracers of ongoing/recently completed star-formation (EW[\ha], \hb, \hd) with the proxy for stellar population age (Dn4000). 

 Of these 210 barred galaxies, we note that 113 host strong bars and 97 host weak bars. All of these galaxies have TW measured pattern speeds; 70 are fast bars (reaching to near co-rotation in their hosts), and 140 are slow (significantly shorter than co-rotation radius). Because the galaxy population is so bimodal in SF properties, even among disc galaxies \citep{Masters2010}, we also separate the barred galaxies into 91 that are star forming (on or above the SF sequence of galaxies) and 119 that are quenching (on the way to be quiescent and below the SF sequence). We create average SF property profiles in sub-samples of fast/strong, slow/strong, fast/weak, and slow/weak for both quiescent and star-forming barred spirals. 

In the SF sample, we find the only statistically significant bar-driven differences in the EW[H$\alpha$] profiles, suggesting the impact of a bar on galaxies actively forming stars is largest on the shortest timescales of SF. Overall, we see that these profiles show a peak of SF in the center, a dip in the bar region, and in many cases, a second peak at or just beyond the bar end. 

As previously observed in the same sample in \cite{geronthesis}, we find evidence that SF galaxies with strong bars show notable ``speed" related differences: slow/strong barred galaxies (green line in Figure \ref{fig:sf_combo}) have the most SF in the bar arms, along with a SF peak at the end of the bar (inside co-rotation). Fast/strong barred galaxies (purple line in Figure \ref{fig:sf_combo}) have significant suppression of SF in the arms, which is matched by an enhancement in Dn4000 in this region, suggesting an older stellar population dominates the light distribution. The SF profile in these galaxies also shows a steep increase towards the bar end at/near co-rotation. 

Repeating this analysis for the quenching/quiescent sample of barred galaxies, we find almost no star formation within any bar radius, which is indicative of inside-out quenching. Even in these quenching galaxies, outside the bar radius, there remains some SF and it is in the galaxies with fast bars (both weak and strong, but particularly notable in the fast-weak bars) that there is the most ongoing recent SF and younger stellar populations beyond the end of the bar. 
 
For the post-starburst tracers (\hb and \hd; but also EW[\ha]), we consider how the radial profile moves points around on 2D plots often used to consider SF properties \citep[e.g.][]{Kauffmann2003, Smethurst2019}. Here, we can plot both SF and quenching samples together, noticing that outskirts of these two samples often -- but not always -- (e.g. the slow-weak subset) have similar properties, while we see significant differences within the bar radius between samples. As with the single radial profiles, in the SF sample we often notice two peaks of recent, or recently completed SF, with the mid-bar regions being suppressed/older. In the quenching samples, the trend is typically a monotonic (although modest) increase in recent SF (and decrease in stellar population age/Dn4000) with radius. 

Seeking to investigate how much of these impacts are global, versus constrained to the bar itself, we compared the 2D radial profiles along the bar, with those in an aperture perpendicular to the bar. We find that, at the bar-ends in the slow/strong samples, all tracers (EW[\ha], \hb, \hd, and Dn4000) are different, suggesting the bar ends are notably more SF active and younger than the non-bar disc region at the bar radius. Beyond the end of the bar, we see this asymmetry persist in the SF subset only. In galaxies with fast/strong bars, we see a difference in only EW[\ha] and Dn4000 at the bar-ends, and in the samples with weak bars (fast or slow), we see no differences at the bar end. This suggests that slow/strong bars create the most bar-localised impacts to SF, while fast/strong bars impact the most recent SF, and weak bars have less local impact.

We build on the findings of \citet{geronthesis}, who showed that slow/strong bars among a ``continuum of bar properties" have the most impact on their host galaxies, and add that this impact is also the most bar-localized (as well as global) in slow/strong bars on short and long timescales. Galaxies with slow/strong bars are both the most likely to be globally quenched among all types of barred galaxies, and when SF is present, show the most significant bar-related asymmetry in SF.
 
 This work demonstrates the complexity of a bar's impact on the SF properties of its host galaxy, which can depend on the global SF properties (presumably linked to gas content), bar strength and bar speed (i.e. the length of the bar relative to the co-rotation radius of the galaxy). The impact of bars as accelerators of disc galaxy quenching is complex and fascinating right across the continuum of bar properties. To fully quantify the role bars play in galaxy evolution will require larger samples of barred galaxies with reliable pattern speed measurements than those to which we currently have access.

\section*{Acknowledgements}

MaNGA is part of Sloan Digital Sky Survey IV. Funding for the Sloan Digital Sky Survey IV was provided by the Alfred P. Sloan Foundation, the U.S. Department of Energy Office of Science, and the Participating Institutions. SDSS-IV acknowledges support and resources from the Center for High Performance Computing at the University of Utah. The SDSS website is www.sdss.org.

This work uses data from proposals GBT16A-095, GBT17A-012, GBT19A-127, GBT20B-033 and GBT21B-130. (HI-MaNGA). The Green Bank Observatory is a facility of the National Science Foundation operated under cooperative agreement by Associated Universities, Inc. 

\section*{Data Availability}

All MaNGA data and associated Value Added Catalogues are available via {\tt https://www.sdss4.org/dr17/manga} or can be access using the {\tt Marvin} python package {\tt https://www.sdss4.org/dr17/manga/marvin/}. A catalog of the bar properties measured using the TW method for the entire TW sample is publicly available at \cite{geron2023_TWsoftware}.
 



\bibliographystyle{mnras}
\bibliography{ref} %



\appendix

\section{Correlations with other galaxy properties}
\begin{figure*}
    \centering
    \begin{tabular}{c c}
        \begin{subfigure}{0.5\textwidth}
            \centering
            \includegraphics[width=\textwidth]{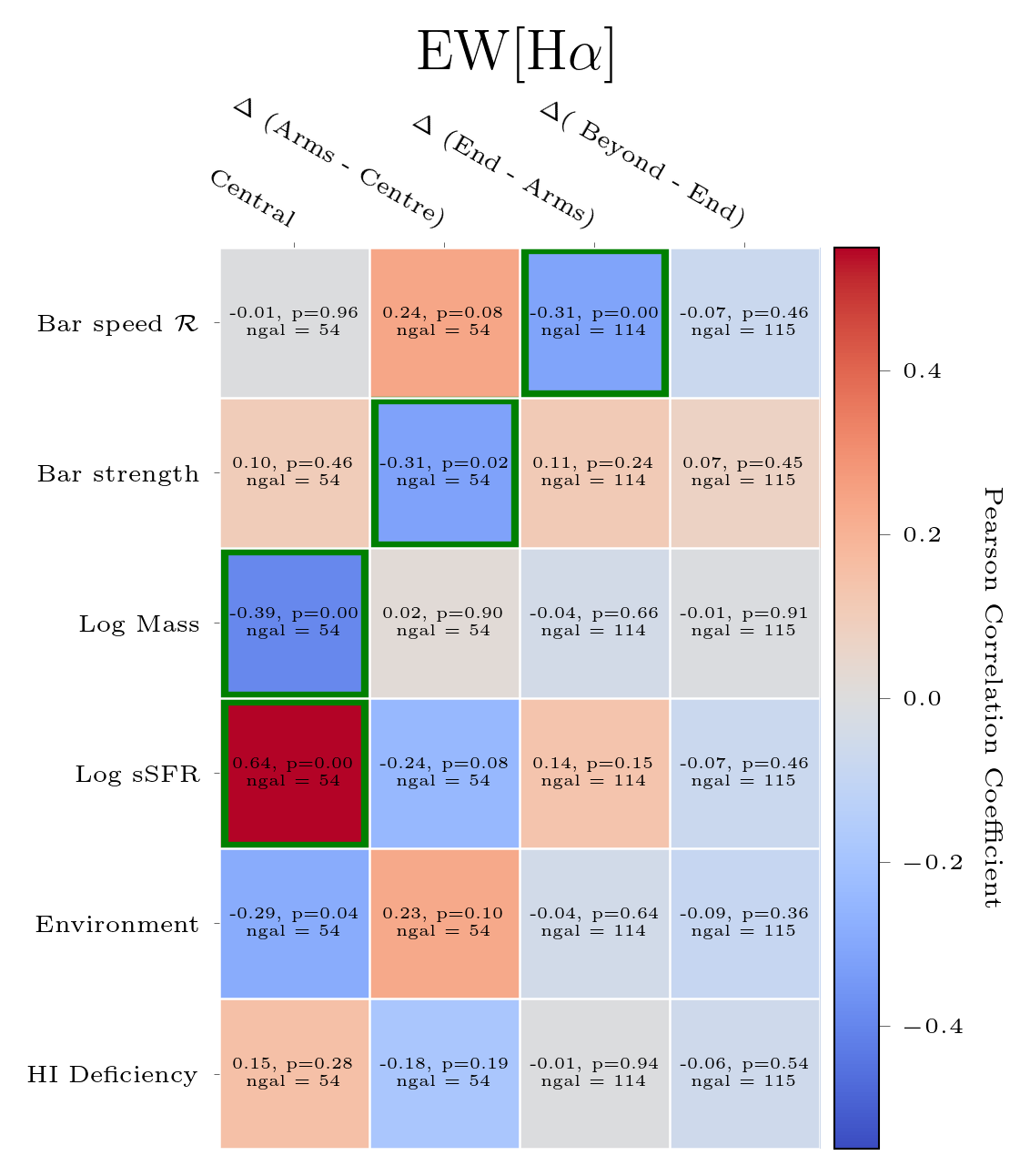}
            \caption{Correlations with measures of EW[H$\alpha$]}
        \end{subfigure} 
        & 
        \begin{subfigure}{0.5\textwidth}
            \centering
            \includegraphics[width=\textwidth]{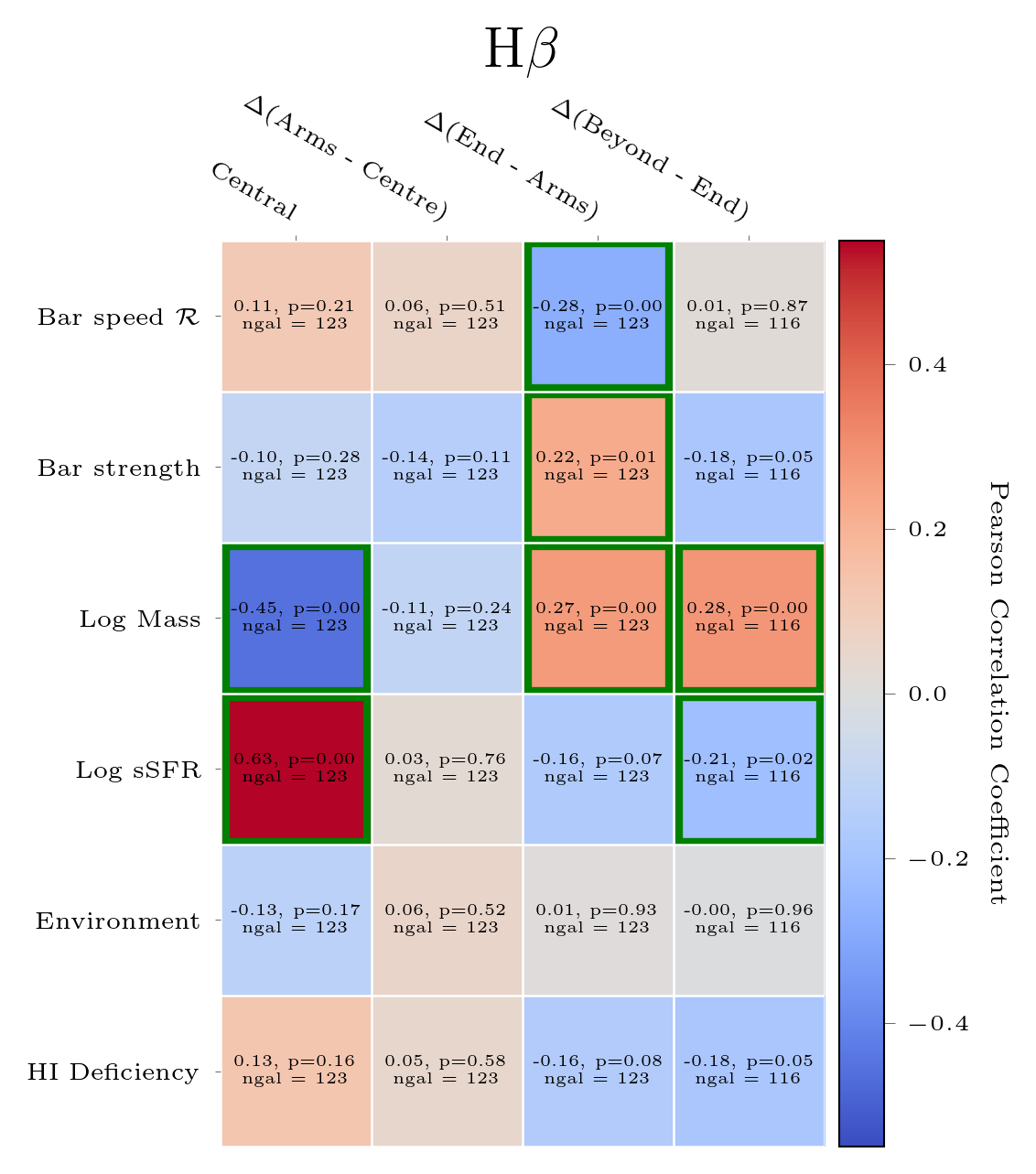}
            \caption{Correlations with measures of \hb}
        \end{subfigure} \\
        
        \begin{subfigure}{0.53\textwidth}
            \centering
            \includegraphics[width=\textwidth]{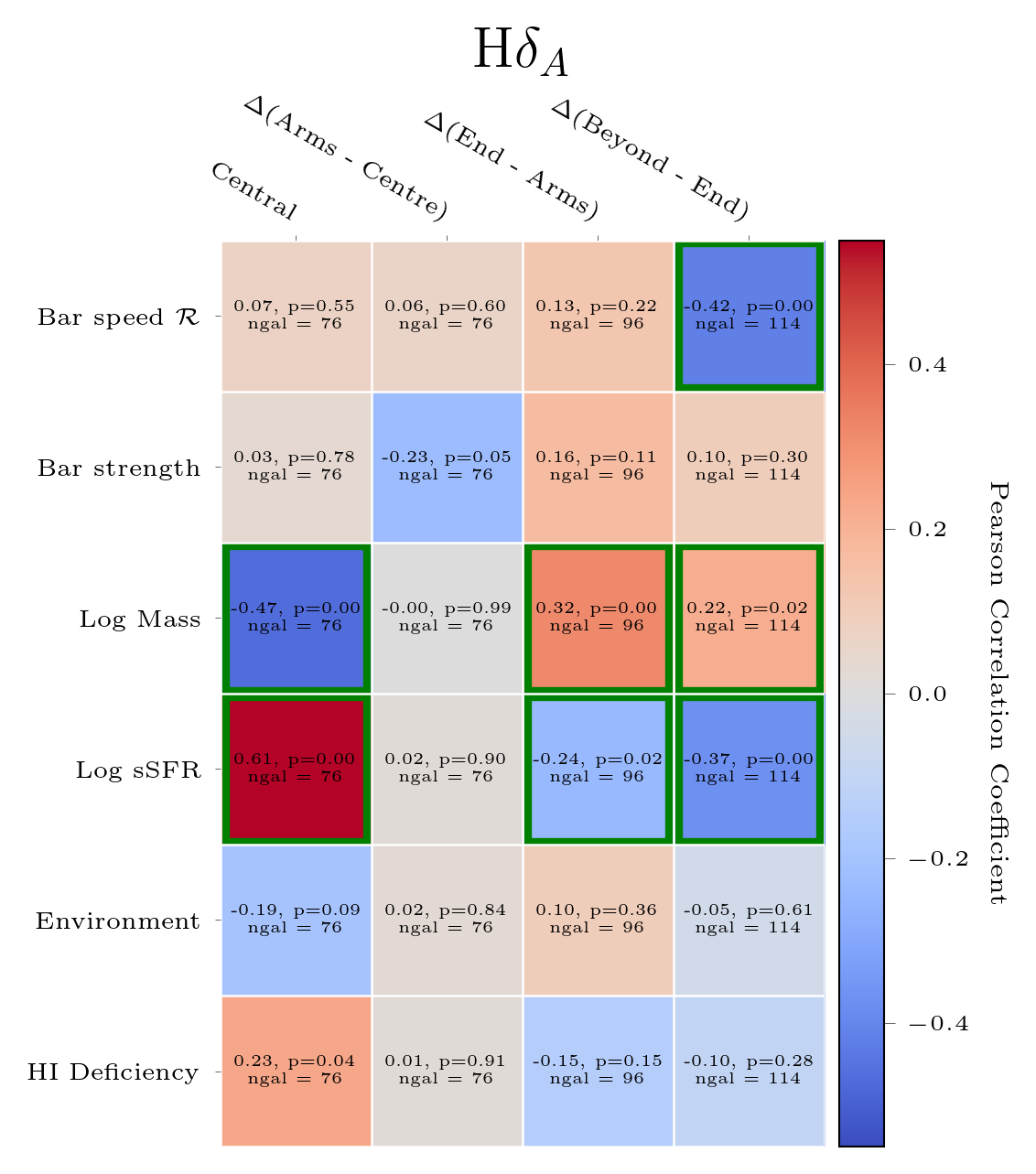}
            \caption{Correlations with measures of \hd}
        \end{subfigure} 
        &  
        \begin{subfigure}{0.53\textwidth}
            \centering
            \includegraphics[width=\textwidth]{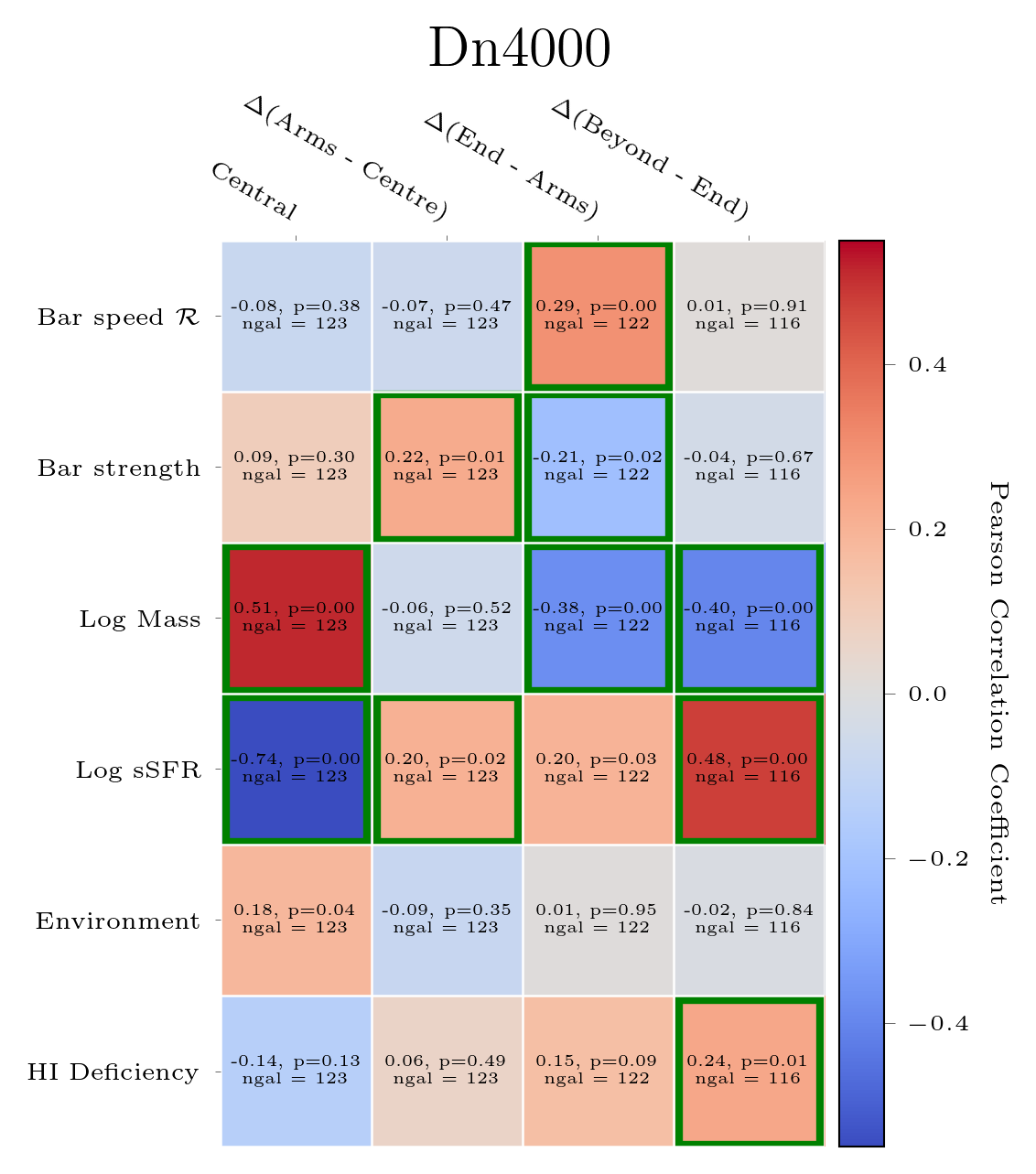}
            \caption{Correlations with measures of Dn4000}
        \end{subfigure} 
    \end{tabular}
    \caption{{Correlation matrices showing the correlation between changes in star-formation/stellar age across four representative regions along the bar (columns in the matrices) and various physical properties of the bar and host galaxy (rows in matrices). Each cell shows (1) the Pearson correlation coefficient between the corresponding measure of star-formation/stellar age and bar/host physical property (which is also used to color-code the cells) (2) the p-value for the correlation and (3) the number of galaxies contributing to the correlation that have measured values for either the star-formation tracers or physical properties in the corresponding region (we remind the reader this may occur for a number of reasons, especially since we remove SF from other contaminating ionization sources)}. }
    \label{fig:correlations}
\end{figure*}

{We present the pairwise correlations between bar-strength, bar-speed, and other physical properties with measures of SF along the barred region in Figure \ref{fig:correlations}. Each cell shows the  $p$-value, number of galaxies, and Pearson correlation coefficient between measures of SF tracers (along columns) and bar/host galaxy physical properties (along rows). Significant correlations ($p<0.03$) are outlined with green boxes.}

\bsp	
\label{lastpage}
\end{document}